\documentclass{jfm}
\usepackage{graphicx}
\usepackage{natbib}
 \usepackage{color}

\let\realverbatim=\verbatim
\let\realendverbatim=\endverbatim
\renewcommand\verbatim{\par\addvspace{6pt plus 2pt minus 1pt}\realverbatim}
\renewcommand\endverbatim{\realendverbatim\addvspace{6pt plus 2pt minus 1pt}}
\makeatletter
\newcommand\verbsize{\@setfontsize\verbsize{10}\@xiipt}
\renewcommand\verbatim@font{\verbsize\normalfont\ttfamily}
\makeatother

\ifCUPmtlplainloaded \else
  \checkfont{eurm10}
  \iffontfound
    \IfFileExists{upmath.sty}
      {\typeout{^^JFound AMS Euler Roman fonts on the system,
                   using the 'upmath' package.^^J}%
       \usepackage{upmath}}
      {\typeout{^^JFound AMS Euler Roman fonts on the system, but you
                   don't seem to have the}%
       \typeout{'upmath' package installed. jfm.cls can take advantage
                 of these fonts,^^Jif you use 'upmath' package.^^J}%
      }
  \else
  \fi
\fi

\ifCUPmtlplainloaded \else
  \checkfont{msam10}
  \iffontfound
    \IfFileExists{amssymb.sty}
      {\typeout{^^JFound AMS Symbol fonts on the system, using the
                'amssymb' package.^^J}%
       \usepackage{amssymb}%
         \let\leq=\leqslant
         \let\geq=\geqslant
      }{}
  \fi
\fi

\ifCUPmtlplainloaded \else
  \IfFileExists{amsbsy.sty}
    {\typeout{^^JFound the 'amsbsy' package on the system, using it.^^J}%
     \usepackage{amsbsy}}
    {\providecommand\boldsymbol[1]{\mbox{\boldmath $##1$}}}
\fi

\providecommand\bnabla{\boldsymbol{\nabla}}
\providecommand\bcdot{\boldsymbol{\cdot}}

\begin{document}


\title[Navier-Stokes transport coefficients of a granular binary mixture]{Modified Sonine approximation for granular binary mixtures}

\author[V. Garz\'{o}, F. Vega Reyes and J. M. Montanero]{V\ls I\ls C\ls E\ls N\ls T\ls E\ns G\ls A\ls R\ls Z\ls \'O$^1$,
F\ls R\ls A\ls N\ls C\ls I\ls S\ls C\ls O\ns V\ls E\ls G\ls A\ns R\ls E\ls Y\ls E\ls S$^1$, \and J\ls O\ls S\ls \'E\ns M\ls A\ls R\ls \'I\ls A\ns M\ls O\ls N\ls T\ls A\ls
N\ls E\ls R\ls O$^2$} \affiliation{$^1$Departamento de F\'{\i}sica, Universidad de
Extremadura, E-06071
Badajoz, Spain,\\ email: vicenteg@unex.es\\[\affilskip]
$^2$ Departamento de Ingenier\'{\i}a Mec\'anica, Energ\'etica y de los Materiales, Universidad de
Extremadura, E-06071 Badajoz, Spain}


\maketitle
\begin{abstract}
We evaluate in this work the hydrodynamic transport coefficients of a granular binary mixture in $d$ dimensions.
In order to eliminate the observed disagreement (for strong dissipation) between computer simulations and
previously calculated theoretical transport coefficients for a monocomponent gas, we obtain explicit expressions of
the seven Navier-Stokes transport coefficients with the use of a new Sonine approach in the Chapman-Enskog theory.
Our new approach consists in replacing, where appropriate in the Chapman-Enskog procedure, the Maxwell-Boltzmann
distribution weight function (used in the standard first Sonine approximation) by the homogeneous cooling state
distribution for each species. The rationale for doing this lies in the fact that, as it is well known, the
non-Maxwellian contributions to the distribution function of the granular mixture become more important in the
range of strong dissipation we are interested in. The form of the transport coefficients is quite common in both
standard and modified Sonine approximations, the distinction appearing in the explicit form of the different
collision frequencies associated with the transport coefficients. Additionally, we numerically solve by means of
the direct simulation Monte Carlo method the inelastic Boltzmann equation to get the diffusion and the shear
viscosity coefficients for two and three dimensions. As in the case of a monocomponent gas,
the modified Sonine approximation improves the estimates of the standard one, showing again the
reliability of this method at strong values of dissipation.
\end{abstract}

\date{\today}
\maketitle

\section{Introduction}
\label{sec1}

The success of the kinetic theory tools in describing granular gases has been widely
recognized \cite[]{G03, BP04}. In particular, the Navier-Stokes (NS) constitutive
equations for the stress tensor and the heat flux of a monocomponent gas have been
derived \cite[]{SG98, BDKS98} by solving the inelastic Boltzmann equation by means of
the Chapman-Enskog (CE) method to first order in the spatial gradients \cite[]{CC70}. As in the
elastic case, the NS transport coefficients are given in terms of the solutions of
linear integral equations \cite[]{BDKS98}, which can be approximately solved to get the
analytical dependence of these coefficients on dissipation. The standard method
consists in approaching the solutions to these integral equations by the
Maxwell-Boltzmann distribution $f_M$ times truncated Sonine polynomial expansions. For
simplicity, usually only the lowest Sonine polynomial (first Sonine approximation) is
retained (we will refer henceforth to this procedure as the {\em standard} Sonine
approximation). In spite of this approximation, the results obtained from this approach
\cite[which formally applies for all values of the coefficient of restitution, see][]{BDKS98}
compare very well with Monte Carlo simulations \cite*[]{BRC99,BRCG00,LBD02,GM02} for
mild degrees of inelasticity, namely, coefficients of restitution $\alpha$ that are
larger than about 0.7. Specifically, while the self-diffusion and shear viscosity
coefficients are well estimated by the first Sonine approximation (even for values of
$\alpha$ smaller than 0.7), the transport coefficients associated with the heat flux
present significant discrepancies with computer simulations \cite[]{BR04,BRMG05} for
high dissipation ($\alpha \lesssim 0.7$). This fact has motivated the search for
alternative methods which provide accurate estimates for the NS transport coefficients
in the complete range of values of $\alpha$. Thus, Noskowicz {\em et al.} (2007) have
used Borel resummation to obtain the distribution function of the homogeneous cooling
state (HCS) and the NS transport coefficients. The method involves the solution of a
system of algebraic equations that must be numerically solved. A different alternative
consists in a slight modification of the Sonine polynomial approach, assuming that the
isotropic part of the first order distribution function is mainly governed by the HCS
distribution $f^{(0)}$ rather than by the Maxwellian distribution $f_M$ \cite[]{L05}
(and this is what we call {\em modified} Sonine approximation). Except for this, this
modified Sonine approximation keeps the usual structure of the standard Sonine
approximation. Comparison with computer simulations \cite*[]{BR04,BRMG05,MSG07} shows
that this modified approximation significantly improves the accuracy of the NS
transport coefficients of the single granular gas at strong dissipation \cite*[]{GSM07},
especially in the case of the heat flux transport coefficients. The improved agreement in the range of high inelasticities is of physical relevance, since it helps to put into proper context the discussion vividly maintained in the field \cite[]{AT06,G03} about the existence (or not) of scale separation in granular gases. Moreover, the range of high inelasticities has growing interest in experimental works \cite*[]{Clerc08,VU08}. Thus, it is important from a fundamental point of view to understand the origin of the discrepancies between theory and simulation for $\alpha \lesssim 0.7$.

Motivated by the good agreement found between theory and simulation for a monocomponent
granular gas, we extend in this work the modified Sonine approach to the case of a
granular binary mixture. Early attempts to obtain the NS transport coefficients were
carried out by means of the CE expansion around Maxwellians at the same temperature for
each species \cite*[]{JM89,Z95,AW98,WA99,SGNT06}. However, as confirmed by computer
simulations \cite*[]{MG02,BT02,DHGD02, PMP02,KT03,WJM03,BRM05,BRM06,SUKSS06} and
experiments \cite*[]{WP02,FM02}, the equipartition assumption is not in general valid in
granular mixtures since it is only close to being fulfilled in the quasielastic limit.
A kinetic theory for granular mixtures at low density which accounts for
nonequipartition effects has been developed in the past few years by
\cite{GD02}. As in the single gas case, the above theory \cite[]{GD02} shows that the
NS transport coefficients of the mixture are given in terms of the solutions of a set
of linear integral equations, which are approximately solved \cite*[]{GD02,GMD06,GM07}
by means of the standard Sonine approximation. Furthermore, this theory has
successfully predicted, as shown by computer simulations
\cite*[]{MG02,DHGD02,BRM05,SUKSS06}, a constant ratio between the granular temperatures
of both species. Here, we revisit the theory of \cite{GD02} and
solve the corresponding linear integral equations defining the NS transport
coefficients by taking the HCS distribution $f_i^{(0)}$ ($i=1,2$) of each species as
the weight function.

As expected, the problem for a binary mixture is much more involved than in the
monocomponent case since not only the number of transport coefficients is larger but
these coefficients are also functions of more parameters (masses, sizes, composition
and three coefficients of restitution). In spite of these technical difficulties,
explicit expressions for the seven relevant transport coefficients of the binary
mixture (the mutual diffusion $D$, the pressure diffusion $D_p$, the thermal diffusion
$D'$, the shear viscosity $\eta$, the Dufour coefficient $D''$, the thermal
conductivity $\lambda$, and the pressure energy coefficient $L$) have been obtained in
terms of the parameter space of the problem. The results show that the modified Sonine
approximation at large inelasticities introduces significant, moderate and slight
corrections to the transport coefficients associated with the heat flux ($D''$,
$\lambda$, $L$), stress tensor ($\eta$) and mass flux ($D$, $D_p$, $D'$), respectively.
In order to assess the degree of accuracy of the modified Sonine approximation, we have
also performed Monte Carlo simulations for the mutual diffusion and the shear viscosity
coefficients and use available simulation data for the heat flux coefficients in the
monodisperse gas case \cite*[]{BR04,BRMG05}. We show that, in the range of strong inelasticity, the modified
Sonine approximation provides better agreement with simulations in all cases, being
this accuracy gain not negligible for viscosity, thermal conductivity, pressure energy
coefficient and Duffour coefficient. This clearly justifies the use of this new approximation in
order to obtain accurate expressions for the NS transport coefficients of the mixture.

An important issue is the applicability of the NS equations for granular binary
mixtures derived here.  The expressions of the seven NS transport coefficients are not
restricted to weak inelasticity and hold for arbitrary values of the degree of
dissipation. In fact, our modified Sonine approximation turns to be more accurate than
the previous one \cite[]{GD02} for very dissipative interactions. However, as already
pointed out by \cite{GMD06}, the NS hydrodynamic equations themselves may or may not be
limited with respect to inelasticity, depending on the particular granular flows of
real materials considered. It is important to recall that the CE method assumes that
the relative changes of the hydrodynamic fields (partial densities, mean flow velocity
and granular temperature) over distances of the order of the characteristic mean free
path of the system are small. In the case of molecular or ordinary gas mixtures, the
strength of the spatial gradients is controlled by external sources (the initial or boundary
conditions, driving forces), and for a great variety of experimental conditions of practical interest the NS hydrodynamic equations describe adequately the hydrodynamics of the system.  Unfortunately, the situation in the case of granular gases is more
complex, since in most problems of practical interest (such as steady states)
the boundary conditions cannot completely control by themselves the spatial gradients of the hydrodynamic fields. This is due to collisional cooling, which sets a minimum value for the relative size of the steady state gradients. An example of this situation corresponds to the well-known
(steady) simple shear flow state \cite*[]{G03,SGD04}. This type of flow can only occur when collisional cooling (which is fixed by the mechanical properties of the particles making up the fluid) is exactly balanced by viscous heating. Unfortunately, this occurs, except for the quasielastic limit, only for very extreme boundary conditions (high shear). The same applies for all steady states for a granular gas heated or sheared from the boundaries, although the behaviour of the gradients is qualitatively different for each type of flow \cite*[for a case by case study, please refer to the work by][]{VU07}. Another example is the case of a granular gas
bounded by two opposite walls at rest and at the same temperature. Recent comparisons
\cite*[]{HGW08} between molecular dynamics simulations and kinetic theory provide
evidence that higher-order effects  (i.e., beyond the NS description) can be measurable for moderate inelasticities.

Nevertheless, in spite of the above cautions, the NS description is still accurate
and appropriate for a wide class of flows. One of them corresponds to small spatial
perturbations of the homogeneous cooling state for an isolated system
\cite*[]{BDKS98,G05}. Both molecular dynamics and Monte Carlo simulations
\cite*[]{BRC99,BRCG00} have quantitatively confirmed the dependence of the  NS
transport coefficients on dissipation (even for strong values of the coefficient of restitution)
and the applicability of the NS hydrodynamics with these transport
coefficients to describe  cluster formation. Another interesting example consists of
the application of the NS hydrodynamics from kinetic theory to characterize the
symmetry breaking as well as the density and temperature profiles in vertical vibrated
gases \cite*[]{BRMG02}. In the case of dense gases, there is evidence of the good
agreement found between the NS coefficients derived from the Enskog kinetic theory
\cite*[]{GD99a} and computer simulations \cite*[]{LBD02,DHGD02,LLC07}. Similar
comparisons between NS hydrodynamics and real experiments of supersonic flow past a
wedge \cite*[]{RBSS02} and Nuclear Magnetic Resonance (NMR) experiments of a system of mustard seeds vibrated
vertically \cite*[]{YHCMW02,HYCMW04} have shown both qualitative and quantitative
agreement. Furthermore, the Navier-Stokes hydrodynamics accounts for a variety of properties of granular flows that are preserved beyond NS order. For instance, the curvature of the temperature profile: the same types of steady profiles predicted by NS hydrodynamics have been observed in DSMC simulations beyond the NS domain, as it is reported in a recent theoretical work where new steady states (exclusive to granular gases) have been found \cite*[]{VU07}. Therefore, the NS equations with the transport coefficients derived here can
be considered still as an useful theory for a wide class of rapid granular flows, although
more limited than for ordinary gases.

The plan of the paper is as follows. In~\S\,\ref{sec2}, the full transport coefficients
of the mixture are given in terms of the solutions of a set of coupled linear integral
equations previously derived by \cite{GD02}. These integral equations are approximately solved by means of the modified
first Sonine approximation in~\S\,\ref{sec3} where explicit forms for the transport
coefficients are provided. Technical details of the calculations carried out here are
given in two Appendices and in a Supplementary Material submitted in the online version of the paper. Next, in~\S\,\ref{sec4} the results obtained for these seven
transport coefficients from the standard and modified Sonine approximations are
compared for several cases. In addition, both theoretical approaches are also compared
with available and new simulation data obtained from numerical solutions of the
Boltzmann equation by using the direct simulation Monte Carlo (DSMC) method
\cite[]{B94} in the cases of the diffusion and shear viscosity coefficients, both for
two- and three-dimensional systems. The paper is closed in~\S\,\ref{sec5} with a
discussion of the results presented in this paper.

\section{Navier-Stokes transport coefficients for a granular binary mixture}
\label{sec2}

We consider a binary mixture composed by smooth inelastic disks ($d=2$) or spheres
($d=3$) of masses $m_{1}$ and $ m_{2} $, and diameters $\sigma _{1}$ and $\sigma _{2}$.
The inelasticity of collisions among all pairs is characterized by three independent
constant coefficients of restitution $\alpha _{11}$, $\alpha _{22}$, and $\alpha
_{12}=\alpha _{21}$, where $\alpha _{ij}\leq 1$ is the coefficient of restitution for
collisions between particles of species $i$ and $j$. In the low-density regime, the
one-particle velocity distribution functions $f_i({\boldsymbol r},{\boldsymbol v},t)$ obey the set of
(inelastic) Boltzmann equations \cite*[]{GS95,BDS97}. When the hydrodynamic gradients
present in the system are weak, the CE method \cite[]{CC70} conveniently adapted to
dissipative dynamics provides a solution to the Boltzmann equation based on an
expansion
\begin{equation}
\label{2.5} f_i=f_i^{(0)}+f_i^{(1)}+\cdots,
\end{equation}
where $f_i^{(0)}$ is the {\em local} version of the homogeneous cooling state (HCS)
\cite[]{GD99b}. Although the exact form of the distribution $f_i^{(0)}$ is not known
(even in the one-component case), an indirect information of the behavior of
$f_i^{(0)}$ is given through its velocity moments. In particular, the deviation of
$f_i^{(0)}$ from its Maxwellian form can be characterized by the fourth cumulant
\begin{equation}
\label{2.5.1} c_i=2\left[\frac{m_i^2}{n_iT_i^2}\frac{1}{d(d+2)}\int\; d{\boldsymbol v}\; V^4
f_i^{(0)}-1\right],
\end{equation}
where ${\boldsymbol V}={\boldsymbol v}-{\boldsymbol U}$ is the peculiar velocity,
\begin{equation}
\label{2.5.2} n_i=\int\; d{\boldsymbol v}\; f_i^{(0)}
\end{equation}
is the number density of species $i$,
\begin{equation}
\label{2.5.3} {\boldsymbol U}=\frac{1}{\rho}\sum_{i=1}^2\;\int\; d{\boldsymbol v}\; m_i{\bf v}\; f_i^{(0)}
\end{equation}
is the mean flow velocity and $\rho=m_1n_1+m_2n_2$ is the total mass density.

The first-order distribution $f_i^{(1)}$ has the form \cite*[]{GD02}
\begin{equation}
f_{i}^{(1)}={\boldsymbol {\cal A}}_{i}\bcdot \bnabla x_{1}+{\boldsymbol {\cal B}}_{i}\bcdot \bnabla p+{\boldsymbol
{\cal C}}_{i}\bcdot \bnabla T+{\cal D}_{i,k\ell }\bnabla _{k}U_{\ell}\;, \label{2.5.1bis}
\end{equation}
where $x_i=n_i/n$ is the mole fraction of species $i$,
\begin{equation}
\label{2.5.4} T=\frac{1}{dn} \sum_{i=1}^2\; \int\;d{\boldsymbol v}\; m_iV^2\; f_i^{(0)}
\end{equation}
is the temperature and $p=nT$ is the pressure. Here, $n=\sum_in_i$ is the total number
density. The coefficients ${\boldsymbol {\cal A}}_{i}$, ${\boldsymbol {\cal B}}_{i}$,
${\boldsymbol {\cal C}}_{i}$, and ${\mathsfbi {\cal D}}_{i}$ are functions of the
peculiar velocity ${\boldsymbol V}={\boldsymbol v}-{\boldsymbol U}$ and the
hydrodynamic fields.

The knowledge of the distributions $f_i^{(1)}$ allows us to determine the NS transport
coefficients. The analysis to obtain them has been previously worked out by Garz\'o
and coworkers \cite[]{GD02,GMD06, GM07}. For the sake of completeness, the final
results will be displayed in this Section. The mass flux ${\bf j}_{1}^{(1)}$, the
pressure tensor $P_{k \ell}^{(1)}$, and the heat flux ${\bf q}^{(1)}$ are given,
respectively, by \cite[]{GD02}
\begin{equation}
{\boldsymbol j}_{1}^{(1)}=-\frac{m_{1}m_{2}n}{\rho } D\bnabla x_{1}-\frac{ \rho }{p}D_{p}\bnabla p-\frac{\rho
}{T}D^{\prime }\bnabla T,\hspace{0.3in}{\boldsymbol j}_{2}^{(1)}=-{\boldsymbol j}_{1}^{(1)},  \label{2.6}
\end{equation}
\begin{equation}
\mathsfbi {P}_{k \ell}^{(1)}=p\;\delta _{k\ell}-\eta \left( \bnabla _{\ell
}U_{k}+\bnabla _{k}U_{\ell}- \frac{2}{d}\delta_{k\ell} {\bnabla \bcdot U}\right),
\label{2.7}
\end{equation}
\begin{equation}
{\boldsymbol q}^{(1)}=-T^{2}D^{\prime \prime }\bnabla x_{1}-L\bnabla p-\lambda \bnabla
T.
\label{2.8}
\end{equation}
It must be noted that the contribution to first order in the gradients of the collisional cooling  vanishes ($\zeta^{(1)}=0$) by symmetry and there is no need to consider here this term \cite[]{BDKS98}. This is due to the fact that the cooling rate is a scalar and so, corrections to first order in gradients can arise only from the divergence of the velocity field. However, as Eq.\ (\ref{2.5.1bis}) shows, there is no contribution to $f_i^{(1)}$ proportional to $\nabla \cdot {\bf U}$ and consequently, $\zeta^{(1)}=0$. This is special to the low density Boltzmann equation and such terms occur at higher densities \cite[]{GD99a,GHD07}.

The transport coefficients appearing in\ (\ref{2.6})--(\ref{2.8}) are the diffusion
coefficient $D$, the pressure diffusion coefficient $D_p$, the thermal diffusion
coefficient $D'$, the shear viscosity $\eta$, the Dufour coefficient $D''$, the
pressure energy coefficient $L$, and the thermal conductivity $\lambda$. These
coefficients are defined as
\begin{equation}
D=-\frac{\rho }{dm_{2}n}\int d{\boldsymbol v}\,{\boldsymbol V}\bcdot {\boldsymbol {\cal A}}_{1}, \label{2.9}
\end{equation}
\begin{equation}
D_{p}=-\frac{m_{1}p}{d\rho}\int d{\boldsymbol v}\,{\boldsymbol V}\bcdot {\boldsymbol {\cal B}}_{1}, \label{2.10}
\end{equation}
\begin{equation}
D^{\prime }=-\frac{m_{1}T}{d\rho }\int d{\boldsymbol v}\,{\boldsymbol V}\bcdot {\boldsymbol {\cal C}}_{1}, \label{2.11}
\end{equation}
\begin{equation}
\eta =-\frac{1}{(d-1)(d+2)}\sum_{i=1}^2\,m_i\,\int d{\boldsymbol v}\, {\boldsymbol V}{\boldsymbol V}:{{\cal D}}_{i},
\label{2.10.1}
\end{equation}
\begin{equation}
D^{\prime \prime }=-\frac{1}{dT^{2}}\sum_{i=1}^2\,\frac{m_i}{2}\,\int d{\boldsymbol
v}\,V^{2}{\boldsymbol V}\bcdot {\boldsymbol {\cal A}}_{i}, \label{2.11.1}
\end{equation}
\begin{equation}
L=-\frac{1}{d}\sum_{i=1}^2\,\frac{m_i}{2}\,\int d{\boldsymbol v}\,V^{2}{\boldsymbol V}
\bcdot \,{\boldsymbol {\cal B}}_{i}, \label{2.12}
\end{equation}
\begin{equation}
\lambda =-\frac{1}{d}\sum_{i=1}^2\,\frac{m_i}{2}\,\int d{\boldsymbol v}\,V^{2}
{\boldsymbol V}\bcdot {\boldsymbol {\cal C}}_{i}. \label{2.13}
\end{equation}
As in the case of elastic collisions \cite[]{CC70}, the unknowns are the
solutions of a set of coupled linear integral equations. Their explicit forms are given in the Supplementary Material file of this article and by Eqs.\ (46)--(49) of \cite{GD02}.

\section{Modified First Sonine approximation}
\label{sec3}

The results presented in the above Section are still exact. However, to get the
explicit dependence of the NS transport coefficients on the parameters of the mixture
(masses, sizes, composition, coefficients of restitution), one has to solve the
integral equations obeying ${\boldsymbol {\cal A}}_{i}$, ${{\cal B}}_{i}$,
${\boldsymbol {\cal C}}_{i}$, and ${\boldsymbol {\cal D}}_{i}$ as well as one has to
know the explicit forms of $f_i^{(0)}$ and of the HCS cooling rate $\zeta^{(0)}$ (see Appendix A and the work by \cite{GD99b} for more details). With respect to the
distribution $f_i^{(0)}$, except in the high velocity region, it is very accurately
estimated by \cite[]{GD99b,MG02}
\begin{equation}
\label{3.1} f_i^{(0)}({\boldsymbol V})=f_{i,M}({\boldsymbol V})\left[1+\frac{c_i}{4}
\left(\theta_i^2V^{*4}-(d+2)\theta_iV^{*2}+\frac{d(d+2)}{4}\right)\right],
\end{equation}
where $c_i$ is defined by\ (\ref{2.5.1}), $V^*=V/v_0$,
$\theta_i=m_i\gamma_i^{-1}\sum_j m_j^{-1}$, and
\begin{equation}
\label{3.1.1} v_0=\sqrt{\frac{2T(m_1+m_2)}{m_1m_2}}
\end{equation}
is a thermal speed. Further, $\gamma_i=T_i/T$ is the temperature ratio and $f_{i,M}$ is
the Maxwellian distribution
\begin{equation}
\label{3.1.1b} f_{i,M}({\boldsymbol V})=n_i\left(\frac{m_i}{2\pi T_i}\right)^{d/2}\exp\left(-\frac{m_iV^2}{2T_i}\right).
\end{equation}
The partial temperatures $T_i$ are defined as
\begin{equation}
\label{3.1.0} T_i=\frac{1}{dn_i}\int\;  d{\boldsymbol v}\; m_iV^2\; f_i^{(0)}.
\end{equation}
The cumulant $c_i$ measures the departure of $f_i^{(0)}$ from $f_{i,M}$. The
temperature ratios $\gamma_i$ along with the coefficients $c_i$ and the cooling rate
$\zeta^{(0)}$ have been determined  for inelastic hard spheres ($d=3$) \cite[]{GD99b}.
These calculations have been extended here in~\S~Appendix \ref{appB} to an arbitrary
number of dimensions.

With respect to the functions $\left\{ {\boldsymbol {\cal A}}_{i}, {\boldsymbol {\cal
B}}_{i}, {\boldsymbol {\cal C}}_{i}, {{\cal D}}_{i} \right\}$, as in the
elastic case, the simplest approximation consists of expanding them in a series of
Sonine polynomials and consider only the leading terms in this expansion. Usually, for
simplicity, the polynomials are defined with respect to a Gaussian weight factor. This
alternative has been previously used \cite[]{GD02,GM07} to get explicit forms for the
NS transport coefficients. The accuracy of these theoretical predictions (based on the
standard first Sonine approximation) over a wide range of inelasticities has been
confirmed by Monte Carlo simulations of the Boltzmann equation in the cases of the
tracer diffusion coefficient \cite[]{GM07,GM04} and the shear viscosity coefficient
\cite[]{GM07,MG03}. However, as said in the Introduction, recent comparisons with
computer simulations for the transport coefficients associated with the heat flux for a
monocomponent gas have shown significant discrepancies for strong inelasticities
\cite[]{BR04,BRMG05,MSG07}. This fact has motivated the search for new approaches, such
as a modified first Sonine approximation \cite[]{GSM07}.

As already pointed out by \cite{GSM07}, one of the possible sources of
discrepancy between the standard first Sonine approximation and computer simulations
for the heat flux transport coefficients could be the emergence of the existing
non-Gaussian features of the zeroth-order distribution function $f_i^{(0)}$. Although
the Maxwellian distribution $f_{i,M}$ is a good approximation to $f_i^{(0)}$ in the
region of thermal velocities relevant to low degree velocity moments (hydrodynamic
quantities, mass flux), quantitative discrepancies between $f_{i,M}$ and $f_i^{(0)}$
are expected to be important when one evaluates higher degree velocity moments, such as
the pressure tensor and the heat flux, specially for strong dissipation. However, the
behavior of the first-order distribution $f_i^{(1)}$ in the standard Sonine
approximation \cite[]{GD02} is mainly governed by the Maxwellian distribution $f_{i,M}$
and not by the HCS distribution $f_i^{(0)}$. A possible way of mitigating the
discrepancies between the standard first Sonine approximation and simulations would be
to incorporate more terms in the Sonine polynomial expansion \cite[]{GM04}, but the
technical difficulties to evaluate these new contributions for general binary mixtures
discard this method. Here, we follow the same route as in our previous work
\cite[]{GSM07} for monocomponent gases and take the distribution $f_i^{(0)}$ instead of
the simple Maxwellian form $f_{i,M}$ as the convenient weight function. In this case,
some care must be taken in the structure of the velocity polynomials chosen to preserve
the solubility conditions of the CE method \cite[]{CC70,GS03}. These conditions are
given by
\begin{equation}
\label{3.2}
 \int d{\boldsymbol v}f_{i}^{(1)}({\boldsymbol v}) =0\;,\
\end{equation}
\begin{equation}
\sum_{i=1}^2\,m_i\,\int d{\boldsymbol v}\,{\boldsymbol v}f_{i}^{(1)}({\boldsymbol v}) ={\boldsymbol 0}\;, \label{3.3}
\end{equation}
\begin{equation}
\sum_{i=1}^2\,\frac{m_{i}}{2}\,\int d{\boldsymbol v}\,V^{2}f_{i}^{(1)}({\boldsymbol v}) =0\;. \label{3.4}
\end{equation}
While the conditions (\ref{3.2}) and (\ref{3.3}) yield the same velocity polynomials as in the standard method, in order to preserve the solubility condition (\ref{3.4}) one has to replace the polynomial
\begin{equation}
\label{3.4.2}
{\boldsymbol S}_i({\boldsymbol V})=\left(\frac{1}{2}m_iV^2-\frac{d+2}{2}T_i\right){\bf V}
\end{equation}
appearing in the standard Sonine polynomial expansion \cite[]{GD02} by the modified polynomial
\begin{eqnarray}
\label{3.4.3}
{\overline {\boldsymbol S}}_i({\boldsymbol V})&=&\left(\frac{1}{2}m_iV^2-\frac{d+2}{2}\left(1+\frac{1}{2}c_i\right)T_i\right){\bf V}\nonumber\\
&=&
{\boldsymbol S}_i({\boldsymbol V})-\frac{d+2}{4}c_iT_i{\bf V}.
\end{eqnarray}
As will be shown later, this replacement gives rise to new contributions to the transport
coefficients associated with the heat flux.

The determination of the NS transport coefficients in the {\em modified} first Sonine
approximation follows similar mathematical steps as the ones previously used in the
{\em standard} first Sonine approximation. More specific technical details on these
calculations are available as a supplement to the online version of this paper or on
request from the authors. Here, we only display the final expressions for the NS
transport coefficients in terms of the parameter space of the problem.

\subsection{Mass flux transport coefficients}

The mass flux contains three transport coefficients: $D$, $D_p$, and $D'$.
Dimensionless forms are defined by
\begin{equation}
\label{3.4.1} D=\frac{\rho T}{m_1m_2\nu_0}D^*,\quad D_p=\frac{p}{\rho \nu_0} D_p^*, \quad D'=\frac{p}{\rho
\nu_0} D^{'*}
\end{equation}
where $\nu _{0}=n\sigma _{12}^{d-1}v_{0}$ is an effective collision frequency and
$\sigma_{12}=(\sigma_1+\sigma_2)/2$. The explicit forms are then
\begin{equation}
D^*=\left( \nu_D-\frac{1}{2}\zeta ^{\ast }\right)^{-1}\left[ \left( \frac{\partial }{\partial x_{1}}x_{1}\gamma
_{1}\right) _{p,T}+\left( \frac{\partial \zeta ^{\ast }}{\partial x_{1}} \right) _{p,T}\left( 1-\frac{\zeta
^{\ast}}{2\nu_D}\right) D_{p}\right] , \label{3.5}
\end{equation}
\begin{equation}
D_{p}^*=x_1\left( \gamma _{1}-\frac{\mu}{x_2+\mu x_1} \right) \left( \nu_D-\frac{3}{2}\zeta ^{\ast}+\frac{\zeta
^{\ast 2}}{2\nu_D}\right) ^{-1}, \label{3.6}
\end{equation}
\begin{equation}
D^{\prime*}=-\frac{\zeta ^{\ast }}{2\nu_D}D_{p}^*. \label{3.7}
\end{equation}
In these equations, $\zeta^*=\zeta^{(0)}/\nu_0$, $\mu=m_1/m_2$ is the mass ratio and
the expression of $\nu_D$ is given by Eq.\ (\ref{a9}) of the Appendix \ref{appA} \footnote{The definitions of collisional frequencies and the calculation of their integrals are given in the accompanying Supplementary Material file.}.

\subsection{Shear viscosity coefficient}

The shear viscosity coefficient $\eta$ is given by
\begin{equation}
\label{3.13} \eta=\frac{p}{\nu_0}\left(x_1T_1^2\eta_1^*+x_2T_2^2\eta_2^*\right),
\end{equation}
where the partial contributions $\eta_i^*$ to the shear viscosity are
\begin{subequations}
\begin{equation}
\label{3.14} \eta_1^*=\frac{2\gamma_2(2\tau_{22}-\zeta^{*})-4\gamma_1\tau_{12}}
{\gamma_1\gamma_2[\zeta^{*2}-2\zeta^{*} (\tau_{11}+\tau_{22})+4(\tau_{11}\tau_{22}-\tau_{12}\tau_{21})]},
\end{equation}
\begin{equation}
\label{3.14.1} \eta_2^*=\frac{2\gamma_1(2\tau_{11}-\zeta^{*})-4\gamma_2\tau_{21}}
{\gamma_1\gamma_2[\zeta^{*2}-2\zeta^{*} (\tau_{11}+\tau_{22})+4(\tau_{11}\tau_{22}-\tau_{12}\tau_{21})]},
\end{equation}
\end{subequations}
where the collision frequencies $\tau_{ij}$ are given by Eqs.\ (\ref{a10})--(\ref{a13}).

\subsection{Heat flux transport coefficients}

The transport coefficients $D''$, $L$, and $\lambda$ associated with the heat flux can be written as
\begin{equation}
D^{\prime \prime }=-\frac{d+2}{2}\frac{n}{(m_{1}+m_{2})\nu_0}\left[ \frac{ x_{1}\gamma _{1}^{3}}{\mu
_{12}}d_{1}^*+\frac{x_{2}\gamma _{2}^{3}}{\mu _{21}}d_{2}^*-\left( \frac{\gamma _{1}}{\mu _{12}}-\frac{\gamma
_{2}}{\mu _{21}}\right) D^{\ast }\right] , \label{3.22}
\end{equation}
\begin{equation}
L=-\frac{d+2}{2}\frac{T}{(m_{1}+m_{2})\nu_0}\left[ \frac{x_{1}\gamma _{1}^{3}}{\mu _{12}}\ell
_{1}^*+\frac{x_{2}\gamma _{2}^{3}}{\mu _{21}}\ell _{2}^*-\left( \frac{\gamma _{1}}{\mu _{12}}-\frac{\gamma
_{2}}{\mu _{21}} \right) D_{p}^{\ast }\right] ,  \label{3.23}
\end{equation}
\begin{equation}
\lambda =-\frac{d+2}{2}\frac{nT}{(m_{1}+m_{2})\nu _{0}}\left[ \frac{ x_{1}\gamma _{1}^{3}}{\mu _{12}}\lambda
_{1}^*+\frac{x_{2}\gamma _{2}^{3}}{\mu_{21}}\lambda _{2}^*-\left( \frac{\gamma _{1}}{\mu _{12}}-\frac{\gamma
_{2}}{ \mu _{21}}\right) D^{\prime }{}^{\ast }\right] , \label{3.24}
\end{equation}
where $\mu_{ij}=m_i/(m_i+m_j)$ and the coefficients $D^{\ast}$, $D_{p}^{\ast }$, and
 $D^{\prime \ast }$ are given by\ Eqs.\ (\ref{3.5})--(\ref{3.7}), respectively. The expressions
 of the (dimensionless) coefficients $d_{i}^{*}$, $\ell
_{i}^*$, and $\lambda _{i}^*$ are
\begin{eqnarray}
\label{3.26} d_1^*&=&\frac{1}{\Lambda}\left\{2\left[2
\nu_{12}Y_2-Y_1(2\nu_{22}-3\zeta^*)\right]\left[\nu_{12}\nu_{21}-\nu_{11}\nu_{22}
+2(\nu_{11}+\nu_{22})\zeta^*-4\zeta^{*2}\right]\right.\nonumber\\
& &+2\left( \frac{\partial \zeta ^{\ast }}{\partial x_{1}}\right)
_{p,T}(Y_3+Y_5)\left[2\nu_{12}\nu_{21}+2\nu_{22}^2-\zeta^*(
7\nu_{22}-6\zeta^{*})\right]\nonumber\\
& & \left. -2\nu_{12}\left( \frac{\partial \zeta ^{\ast }}{\partial
x_{1}}\right)_{p,T}(Y_4+Y_6)\left(2\nu_{11}+2\nu_{22}-7\zeta^*\right)\right\},
\end{eqnarray}
\begin{eqnarray}
\label{3.27} \ell_1^*&=&\frac{1}{\Lambda}\left\{-2Y_3\left[2
(\nu_{12}\nu_{21}-\nu_{11}\nu_{22})\nu_{22}+\zeta^*(7\nu_{11}\nu_{22}-5\nu_{12}\nu_{21}+2\nu_{22}^2
\right.\right.\nonumber\\
& &\left. -6\nu_{11}\zeta^*-7\nu_{22}\zeta^*+6\zeta^{*2})\right]+
2Y_4\nu_{12}\left[2\nu_{12}\nu_{21}-2\nu_{11}\nu_{22}+2\zeta^*(\nu_{11}+\nu_{22})
-\zeta^{*2}\right]\nonumber\\
& +& \left.2Y_5\zeta^*\left[2\nu_{12}\nu_{21}+\nu_{22}(2\nu_{22}-7\zeta^*)+6\zeta^{*2}\right]
-2\nu_{12}\zeta^*Y_6\left[2(\nu_{11}+\nu_{22})-7\zeta^*\right] \right\},
\end{eqnarray}
\begin{eqnarray}
\label{3.28} \lambda_1^*&=&\frac{1}{\Lambda}\left\{-Y_3\zeta^*\left[2
\nu_{12}\nu_{21}+\nu_{22}(2\nu_{22}-7\zeta^*)+6\zeta^{*2}\right]+\nu_{12}\zeta^*
Y_4\left[2(\nu_{11}+\nu_{22})-7\zeta^*\right] \right. \nonumber\\
& &-Y_5\left[4\nu_{12}\nu_{21}(\nu_{22}-\zeta^*)+2\nu_{22}^2(5\zeta^*-2\nu_{11})+2\nu_{11}
(7\nu_{22}\zeta^*-6\zeta^{*2})+5\zeta^{*2}(6\zeta^*-7\nu_{22})\right]\nonumber\\
& & \left. +\nu_{12}Y_6\left[4\nu_{12}\nu_{21}+2\nu_{11}(5\zeta^*-2\nu_{22})+\zeta^*(10\nu_{22}-
23\zeta^*)\right]\right\}.
\end{eqnarray}
Here, the Y's are defined by\ (\ref{3.30})--(\ref{3.35}),
\begin{equation}
\label{3.29} \Lambda=\left[4(\nu_{12}\nu_{21}-\nu_{11}\nu_{22})+6\zeta^*(\nu_{11}+\nu_{22})-9\zeta^{*2}\right]
\left[\nu_{12}\nu_{21}-\nu_{11}\nu_{22}+2\zeta^*(\nu_{11}+\nu_{22})-4\zeta^{*2}\right],
\end{equation}
and the (reduced) collision frequencies $\nu_{ij}$ are given by Eqs.\ (\ref{a14}) and
(\ref{a15}). The corresponding expressions for $d_2''$, $\ell_2$, $\lambda_2$,
$\nu_{22}$, and $\nu_{21}$ can be deduced from\ (\ref{3.26})--(\ref{3.28}) by
interchanging $1\leftrightarrow 2$. With these results, the coefficients $D''$, $L$,
and $\lambda$ can be explicitly obtained from\ (\ref{3.22})--(\ref{3.24}). In\
(\ref{3.26})--(\ref{3.28}) it is understood that the coefficients $D^*$, $D_p^*$, and
$D^{'*}$ are given by\ (\ref{3.5})--(\ref{3.7}), respectively. Of course, our results
show that $D^{''}$ is antisymmetric with respect to the change $1\leftrightarrow 2$
while $L$ and $\lambda$ are symmetric. Consequently, in the case of mechanically
equivalent particles ($m_1=m_2$, $\sigma_1=\sigma_2$, $\alpha_{ij}=\alpha$), the
coefficient $D^{''}=0$.

The expressions for the NS transport coefficients derived here for a granular binary
mixture reduce to those previously obtained \cite[]{GM07} when one takes Maxwellians
distributions for the reference homogeneous cooling state $f_i^{(0)}$ ($c_1=c_2=0$). In
addition, for mechanically equivalent particles, the results obtained by Garz\'o {\em
et al.} (2007c)  for a single gas by using the modified Sonine method are also
recovered. This confirms the self-consistency of the results reported in this paper.

\begin{figure}
\begin{center}
\begin{tabular}{lr}
\resizebox{6.5cm}{!}{\includegraphics{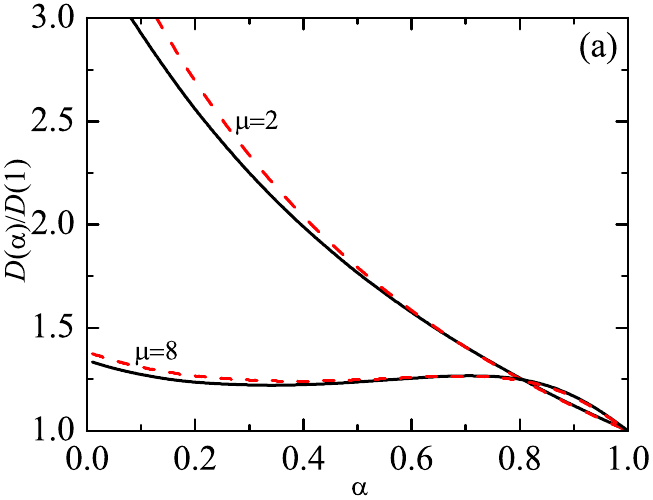}}&\resizebox{6.5cm}{!}{\includegraphics{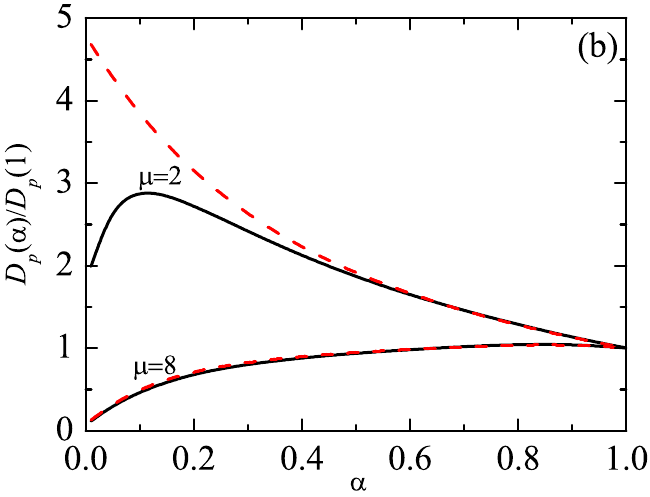}}
\end{tabular}
\resizebox{6.5cm}{!}{\includegraphics{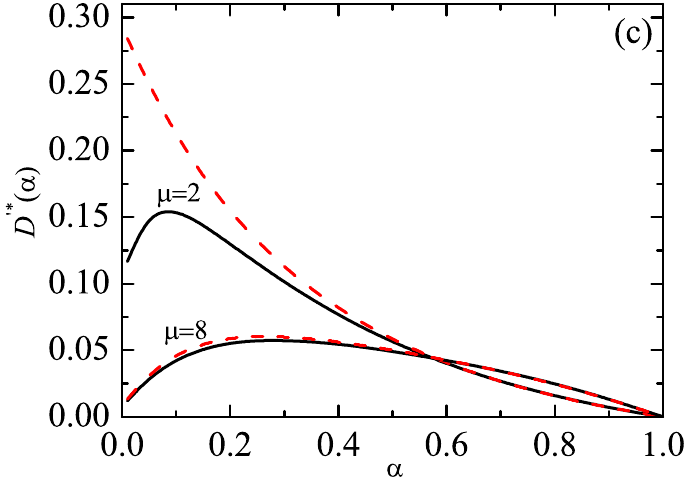}}\\
\end{center}
\caption{Plot of the reduced coefficients $D(\alpha)/D(1)$ (a),
$D_p(\alpha)/D_p(1)$ (b), and $D^{'*}(\alpha)$ (c) as functions of the  coefficient of
restitution $\alpha$ for hard spheres with $x_1=\frac{1}{2}$, $\sigma_1=\sigma_2$ and
two different values of the mass ratio $\mu=m_1/m_2$. The solid lines correspond to the
results obtained from the modified first Sonine approximation while the dashed lines
refer to the results obtained from the standard first Sonine approximation.
\label{fig1}}
\end{figure}

\begin{figure}
\begin{center}
\includegraphics[width=.55\columnwidth]{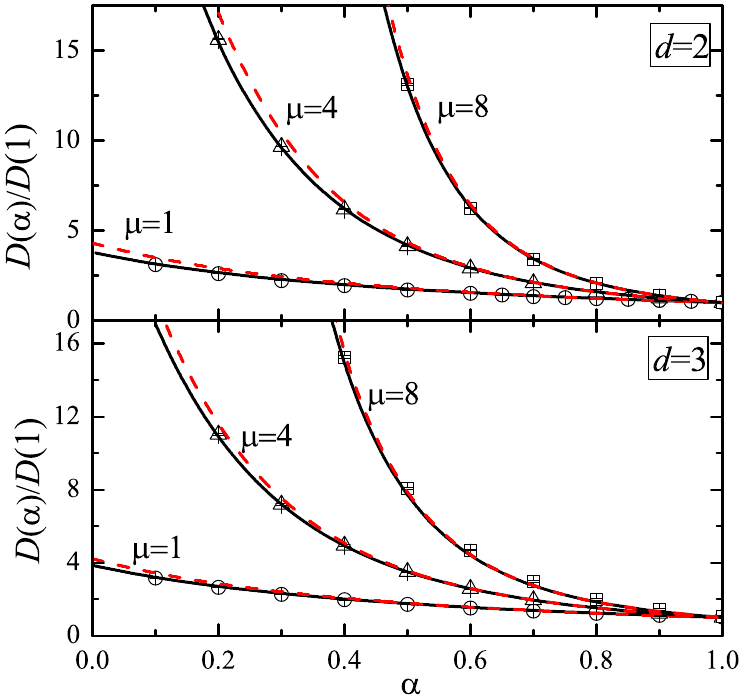}
\end{center}
\caption{Plot of the reduced diffusion coefficient
$D(\alpha)/D(1)$ as a function of the  coefficient of
restitution $\alpha$ for hard disks (top panel) and hard spheres (bottom panel) in the tracer limit ($x_1\to 0)$
when the tagged particle has the same mass density as the particles of the gas ($\mu=\omega^d$).
Three different values of the
mass ratio have been considered: $\mu=1$, $\mu=4$ and $\mu=8$.
The solid and dashed lines represent the modified and standard first Sonine approximations, respectively. The symbols
are DSMC results obtained from the mean square displacement of the tagged particle. The DSMC results
correspond to $\mu=1$ (circles), $\mu=4$ (triangles), and $\mu=8$ (squares).
\label{fig2}}
\end{figure}

\section{Comparison with the standard first Sonine approximation and with Monte Carlo simulations}
\label{sec4}

The expressions for the transport coefficients derived in the previous Section depend
on many parameters: $\{x_1, m_1/m_2, \sigma_1/\sigma_2, \alpha_{11}, \alpha_{22},
\alpha_{12}\}$. Obviously, this complexity exists in the elastic limit as well, so that
the primary new feature is the dependence of the transport coefficients on dissipation.
Thus, to show more clearly the influence of inelasticity in collisions on transport, we
normalize the transport coefficients with respect to their values in the elastic limit.
Also, for simplicity, we take the simplest case of common coefficient of restitution
($\alpha_{11}=\alpha_{22}=\alpha_{12}\equiv \alpha$). This reduces the parameter space
to four quantities: $\{x_1, m_1/m_2, \sigma_1/\sigma_2, \alpha\}$.

\begin{figure}
\begin{center}
\includegraphics[width=.55\columnwidth]{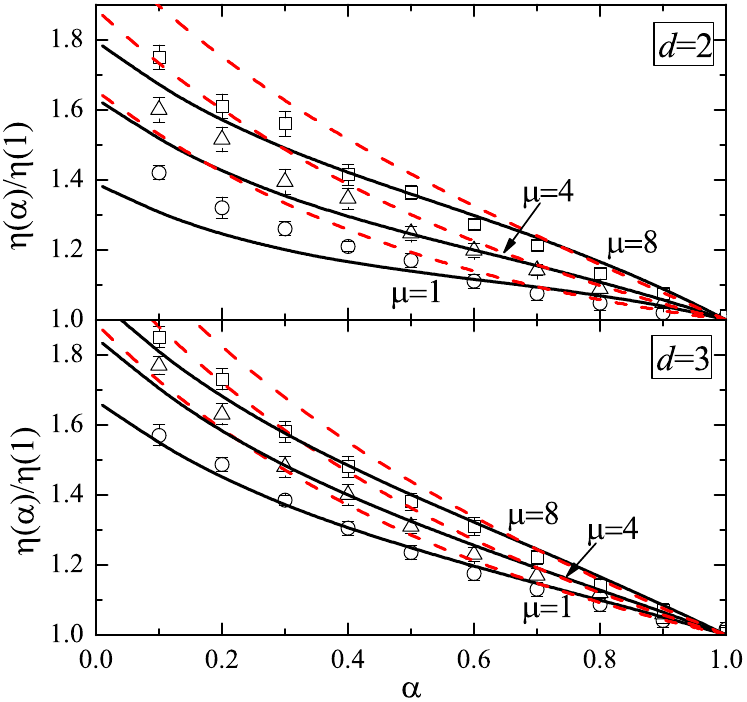}
\end{center}
\caption{Plot of the reduced shear viscosity coefficient
$\eta(\alpha)/\eta(1)$ as a function of the  coefficient of restitution $\alpha$ for
hard disks (top panel) and hard spheres (bottom panel) for an equimolar mixture
($x_1=\frac{1}{2})$ constituted by particles of the same mass density ($\mu=\omega^d$).
Three different values of the mass ratio have been considered: $\mu=1$, $\mu=4$ and
$\mu=8$. The solid and dashed lines represent the modified and standard first Sonine
approximations, respectively. The DSMC results correspond to $\mu=1$ (circles), $\mu=4$
(triangles), and $\mu=8$ (squares). \label{fig3}}
\end{figure}

Let us start with the coefficients $D$, $D_p$ and $D'$ associated with the mass flux.
In Fig.\ \ref{fig1} the reduced coefficients $D(\alpha)/D(1)$, $D_p(\alpha)/D_p(1)$ and
$D^{'*}$, defined by\ (\ref{3.7}), are plotted as functions of the coefficient of
restitution $\alpha$ for an equimolar mixture $x_1=\frac{1}{2}$ of hard spheres ($d=3$)
with the same size ratio $\omega\equiv\sigma_1/\sigma_2=1$ and three different values
of the mass ratio $\mu\equiv m_1/m_2$. Here, $D(1)$ and $D_p(1)$ refer to the elastic
values of $D$ and $D_p$, respectively. The coefficient $D^{'}$ has not been reduced
with respect to its elastic value since $D'=0$ for $\alpha=1$. We observe that both
Sonine approximations lead to quite identical results, except for quite extreme values
of dissipation where both approaches present some discrepancies. This is especially
important in the case $\mu=2$ where the disagreement is, for instance, about 15\% and
32\% for the coefficient $D_p$ at $\alpha=0.1$ and 0.2, respectively. In order to test
the accuracy of both Sonine approximations in the case of the diffusion coefficients,
the dependence of the (reduced) diffusion coefficient $D(\alpha)/D(1)$ on $\alpha$ is
plotted in Fig.\ \ref{fig2} in the tracer limit ($x_1\to 0$) for two different cases
when the tagged particle has the same mass density as the particles of the gas (i.e.,
$\mu=\omega^d$). The theoretical predictions of both Sonine approximations are compared
with available \cite[]{GM04,GM07} and new Monte Carlo simulations \cite[using the DSMC
method,][]{B94}. Here, the tracer diffusion coefficient has been measured in computer
simulations from the mean square displacement of a tagged particle in the HCS
\cite[]{GM04}. It is apparent that both Sonine approximations provide a general good
agrement with simulation data. However, the standard approximation slightly
overestimates the diffusion coefficient at high inelasticity, this effect being
corrected by the modified approximation.

\begin{figure}
\begin{center}
\begin{tabular}{lr}
\resizebox{6.5cm}{!}{\includegraphics{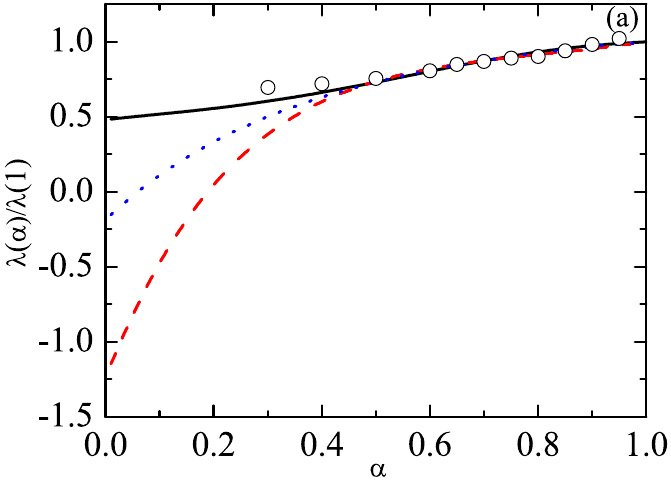}}&\resizebox{6.5cm}{!}{\includegraphics{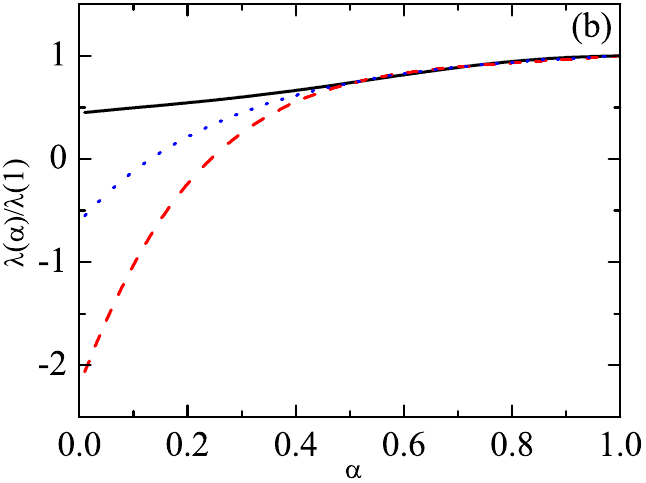}}
\end{tabular}
\resizebox{6.5cm}{!}{\includegraphics{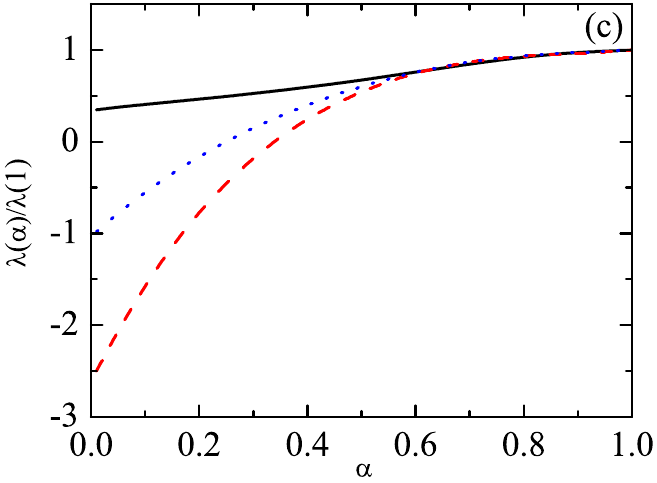}}
\end{center}
\caption{Plot of the reduced thermal conductivity coefficient
$\lambda(\alpha)/\lambda(1)$ for hard spheres with $x_1=\frac{1}{2}$,
$\sigma_1=\sigma_2$ and three different values of the mass ratio $\mu$: $\mu=1$ (a),
$\mu=2$ (b), and $\mu=8$ (c). The solid lines correspond to the results obtained from
the modified first Sonine approximation, the dashed lines refer to the results obtained
from the standard first Sonine approximation while the dotted lines are the results
derived by neglecting non-Gaussian corrections to the HCS distributions. The symbols in
the panel (a) are DSMC results obtained from the Green-Kubo relations for a
monocomponent gas \cite[]{BR04}. \label{fig4}}
\end{figure}

\begin{figure}
\begin{center}
\begin{tabular}{lr}
\resizebox{6.5cm}{!}{\includegraphics{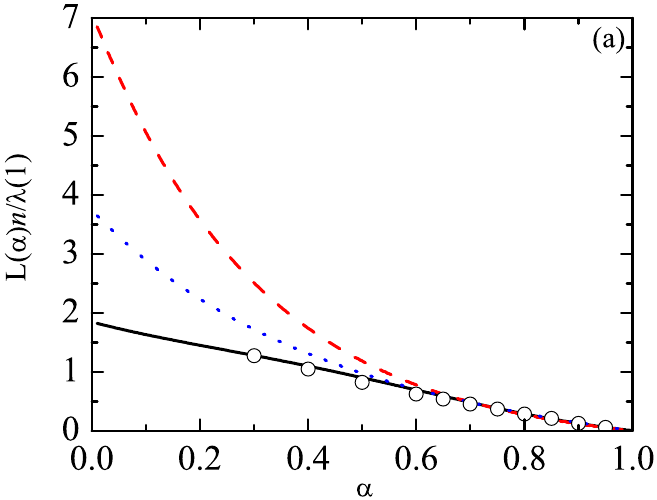}}&\resizebox{6.5cm}{!}{\includegraphics{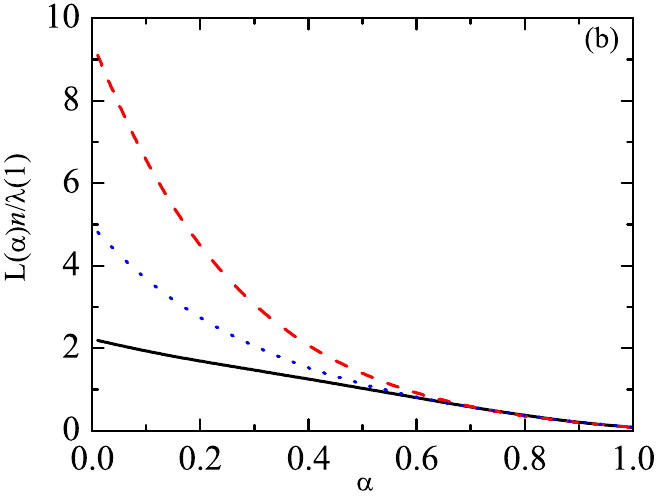}}
\end{tabular}
\resizebox{6.5cm}{!}{\includegraphics{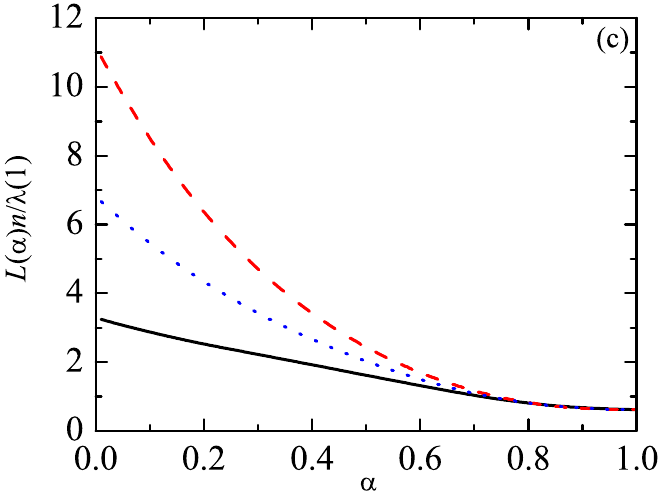}}
\end{center}
\caption{Plot of the reduced pressure energy coefficient
$L(\alpha)n/\lambda(1)$ for hard spheres with $x_1=\frac{1}{2}$, $\sigma_1=\sigma_2$
and three different values of the mass ratio $\mu$: $\mu=1$ (a), $\mu=2$ (b), and
$\mu=8$ (c). The solid lines correspond to the results obtained from the modified first
Sonine approximation, the dashed lines refer to the results obtained from the standard
first Sonine approximation while the dotted lines are the results derived by neglecting
non-Gaussian corrections to the HCS distributions. The symbols in the panel (a) are
DSMC results obtained from the Green-Kubo relations for a monocomponent gas
\cite[]{BR04}. \label{fig5}}
\end{figure}

\begin{figure}
\begin{center}
\begin{tabular}{lr}
\resizebox{6.5cm}{!}{\includegraphics{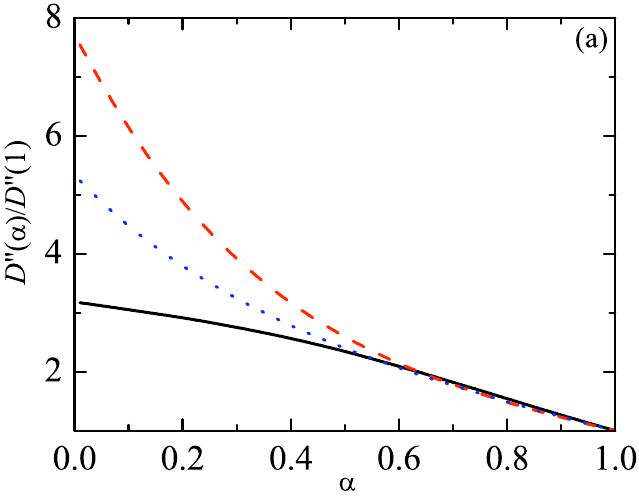}}&\resizebox{6.5cm}{!}{\includegraphics{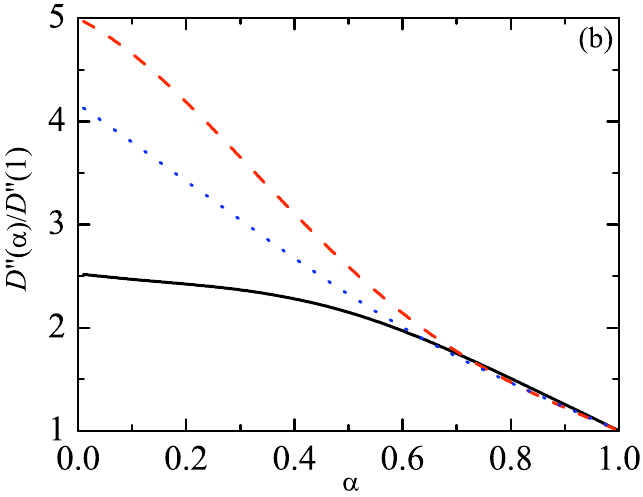}}
\end{tabular}
\resizebox{6.5cm}{!}{\includegraphics{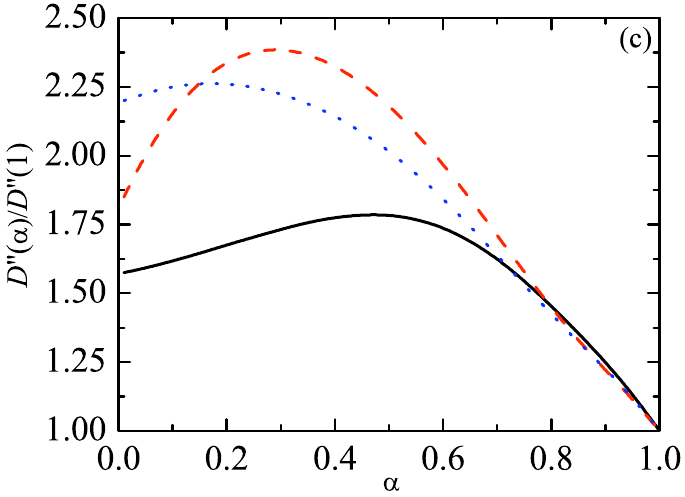}}
\end{center}
\caption{Plot of the reduced Duffour coefficient $D''(\alpha)/D(1)$
for hard spheres with $x_1=\frac{1}{2}$, $\sigma_1=\sigma_2$ and three different values of
the mass ratio $\mu$: $\mu=2$ (a), $\mu=4$ (b), and $\mu=8$ (c).
The solid lines correspond to the results obtained from the modified first Sonine approximation,
the dashed lines refer to the results obtained from the standard first Sonine approximation while
the dotted lines are the results derived by neglecting non-Gaussian
corrections to the HCS distributions.
\label{fig6}}
\end{figure}

The shear viscosity $\eta$ is perhaps the most widely studied transport coefficient in
granular fluids. Here, this coefficient has been measured in simulations by means of a
new method proposed by the authors \cite[]{MSG05}. This method is based on the simple
shear flow state modified by the introduction of (i) a deterministic nonconservative
force (Gaussian thermostat) that compensates for the collisional cooling and (ii) a
stochastic process. While the Gaussian external force allows the granular mixture to
reach a Newtonian regime where the (true) Navier-Stokes shear viscosity can be
identified, the stochastic process is introduced to reproduce the conditions appearing
in the CE solution to Navier-Stokes order. More details on this procedure can be found
in Montanero {\em et al.} (2005). The simulation data obtained from this method along
with both Sonine approximations are presented in Fig.\ \ref{fig3} for disks ($d=2$) and
spheres ($d=3$). We have considered again mixtures constituted by particles of the same
mass density. The simulation data corresponding to $d=2$ for $\alpha \geq 0.5$ were
reported by \cite{GM07} while those corresponding to $d=3$ and $d=2$ for $\alpha \leq
0.5$ have been obtained in this work. As in the case of a single gas \cite[]{GSM07}, we
observe that up to $\alpha \simeq 0.6$ the simulation data agree quite well with both
theories. On the other hand, for higher inelasticities and in the physical
three-dimensional case, the standard first Sonine approximation overestimates the
shear viscosity while the modified first Sonine approximation compares well with
computer simulations, even for low values of $\alpha$. We also observe that the improvement of
the modified Sonine method over the standard one is less clear for hard disks
($d=2$) since the numerical data lie systematically between both theoretical
approaches. However, even in this case the simulation results are closer to the modified ones than the standard ones.
Therefore, according to the comparison carried out at the level of the coefficients $D$
and $\eta$, we can conclude that while the standard Sonine approximation does quite
good a job for not strong values of dissipation, it is fair to say that the modified
Sonine approximation is still better (especially for hard spheres) since is able
to agree well with computer simulations in the full range of values of dissipation
explored.

Let us consider finally the heat flux. As said in the Introduction, recent studies for
a monocomponent gas \cite[]{BR04,BRMG05,MSG07} have shown that the standard first
Sonine approximation dramatically overestimates the $\alpha$-dependence of the
transport coefficients associated with the heat flux for high dissipation ($\alpha
\lesssim 0.7$). This is the main reason why new alternative approximations
\cite[]{NBSG07} have been proposed. However, in contrast to the single gas case, the
lack of available simulation data for granular mixtures for the coefficients $D''$, $L$
and $\lambda$ precludes a comparison between both Sonine approximations and computer
simulations. Figures \ref{fig4}, \ref{fig5} and \ref{fig6} show the $\alpha$-dependence
of the reduced transport coefficients $\lambda(\alpha)/\lambda(1)$,
$L(\alpha)n/\lambda(1)$, and $D''(\alpha)/D''(1)$, respectively, for $d=3$,
$x_1=\frac{1}{2}$, $\omega=1$ and different values of the mass ratio. Also for
comparison we show the theoretical results obtained by neglecting non-Gaussian
corrections to the HCS distributions $f_i^{(0)}$ [i.e., the cumulants given in\
(\ref{2.5.1}) are neglected: $c_i=0$]. Simulation results reported by Brey \&
Ruiz-Montero (2004) for a monocomponent granular gas are also included in the case
$\mu=1$, showing that the modified Sonine appoximation agrees with simulation data
significantly better than the standard Sonine approximation, especially in the case of
the pressure energy coefficient $L$. As expected, the standard and modified Sonine
theories differ significantly for strong inelasticity, especially  for mechanically
{\em different} particles  ($\mu\neq 1$). These discrepancies clearly justify the use
of the modified Sonine approximation instead of the standard one to compute the
dependence of the heat flux transport coefficients on dissipation. In addition, the
standard approach leads to unphysical (negative) values for the thermal conductivity
coefficient $\lambda$ for low values of $\alpha$, which is corrected by our modified Sonine approximation. This is of relevance, since it helps to clarify that the origin of unphysical values for the thermal conductivity at very high inelasticity is due to the mathematical accuracy  of the Sonine approximation used and not to a more fundamental reason; an inherent lack of scale separation, for instance. Note also that the simple theoretical
results \cite[]{GM07} obtained by using Maxwellian forms for $f_i^{(0)}$ are even
closer to the modified ones than those predicted by the standard approximation. The good agreement found in general between simulation (which yield the "actual" values of the transport coefficients) and the modified Sonine approximation is a first evidence that the Chapman-Enskog method may be safely extended down to very low values of the coefficient of normal restitution $\alpha$.

\section{Discussion}
\label{sec5}

A modified Sonine approximation recently proposed \cite[]{GSM07} for monocomponent
systems has been extended to granular mixtures in this paper. This theory has been
mainly motivated by the disagreement found at high dissipation
\cite[]{BR04,BRMG05,MSG07} between the simulation data for the heat flux transport
coefficients and the expressions derived from the standard first Sonine approximation
for a single gas. As we have shown, important discrepancies between computer
simulations and the standard Sonine approximation appear only in the region of strong
inelasticity ($\alpha\lesssim 0.7$). Thus, it could be argued that the search for new theoretical methods may
admittedly be more a mathematical than a physical endeavor. However, it is still
physically relevant to propose methods \cite[]{NBSG07,GSM07} that produce accurate
results for the transport coefficients in the complete range of possible values of the
coefficients of restitution. As in the case of a monocomponent gas \cite[]{GSM07}, in
the modified Sonine approximation the weight function for the unknowns ${\boldsymbol
{\cal A}}_{i}$, ${\boldsymbol {\cal B}}_{i}$, ${\boldsymbol {\cal C}}_{i}$, and
${\boldsymbol {\cal D}}_{i}$ defining the first order distribution $f_i^{(1)}$ is not
longer the Maxwell-Boltzmann distribution $f_{i,M}$ but the HCS distribution
$f_i^{(0)}$. Moreover, in order to preserve the solubility conditions
(\ref{3.2})--(\ref{3.4}), the polynomial ${\boldsymbol S}_i({\boldsymbol V})$, defined by\
(\ref{3.4.2}), appearing in the standard Sonine polynomial expansion must be replaced by
the modified polynomial ${\overline {\boldsymbol S}}_i({\boldsymbol V})$, defined by\
(\ref{3.4.3}). The idea behind the modified method is that the deviation of $f_i^{(0)}$
from $f_{i,M}$ has an important influence on the NS distribution $f_i^{(1)}$. In this context, it
is expected that the rate of convergence of the Sonine polynomial expansion is accelerated when
$f_i^{(0)}$ rather than $f_{i,M}$ is used as weight function.

The extension of the modified Sonine approach to mixtures is not easy at all since it
involves the computation of new complex collision integrals. However, the structure of
the NS transport coefficients is quite similar in the standard and modified
approximations \cite[]{GD02} since the distinction between both approximations occurs
essentially in the $\alpha$-dependence of the characteristic collision frequencies of
the transport coefficients $\nu_D$, $\tau_{ij}$, $\omega_{ij}$ and $\nu_{ij}$. The
forms of the NS coefficients are given by equations\ (\ref{3.4.1})--(\ref{3.7}) for the
mass flux transport coefficients $D$, $D_p$, and $D'$, equations
(\ref{3.13})--(\ref{3.14.1}) for the shear viscosity $\eta$ and equations\
(\ref{3.22})--(\ref{3.28}) for the heat flux transport coefficients $\lambda$, $L$, and
$D''$. All the above expressions have the power to be explicit, namely, they are explicitly given in
terms of the parameters of the mixture. A {\sc Mathematica} code providing the NS
transport coefficients under arbitrary values of composition, masses, sizes, and
coefficients of restitution can be downloaded from our website
\footnote{http://www.unex.es/fisteor/vicente/granular\_files.html/ }. The fact that our
theory does not involve numerical solutions allows us to evaluate the transport
coefficients within very short computing times. For instance, the code using our
theoretical expressions for the NS transport coefficients takes just of the order of 5
seconds in a standard personal computer to produce a graph similar to those in Fig.\
\ref{fig6} of the most complicated heat flux transport coefficients
\footnote{Calculations were performed in a \textsc{PC} equipped with an x86 64-bit
Intel\circledR~processor and a \textsc{RAM} memory of 2 \textsc{GB}}. These results
contrast with the method devised by Noskowicz {\em et al.} (2007) for a mocomponent gas
where a system of algebraic equations must be numerically solved by employing the power
of symbolic processors. This is perhaps another new added value of the method developed
here for granular mixtures.

In order to check the accuracy of the modified Sonine approximation, computer
simulations based on the DSMC method have been carried out in the cases of the tracer
diffusion coefficient $D$ and the shear viscosity coefficient $\eta$. Although some
simulation data for these coefficients were previously reported by the authors
\cite[]{GM04,MSG05}, in this paper we extend those simulations to very low values of
the coefficient of restitution (values of $\alpha$ typically larger than 0.1). This
allows one to make a careful comparison between the modified and standard
approximations with computer simulations. As expected, the discrepancies between
simulation and the standard first Sonine estimates for $D$ and $\eta$ are partially
corrected by the modified approximation, showing again the reliability of such approach
for very strong values of dissipation. Unfortunatelly, the lack of available simulation
data for the coefficients $\lambda$, $L$ and $D''$ corresponding to the heat flux
precludes a comparison between theory and simulation for these transport coefficients,
except in the single gas case \cite[]{BR04,BRMG05} where Figs. \ref{fig5} and
\ref{fig6} show again the superiority of the modified Sonine approximation. We expect
the present results for binary granular mixtures stimulates the performance of such
simulations to assess the degree of accuracy of the modified Sonine method for the heat
flux transport coefficients.

Hydrodynamic theories based on kinetic theory tools have been successful at predicting not only rapid
granular flows (where the role of the interstitial fluid is assumed negligible), but have also been incorporated
into models of high-velocity, gas-solid systems. These kinetic theory calculations are now standard features of commercial and research codes, such as Fluent$^{\circledR}$ and MFIX (http://www.mfix.org/). These codes rely upon accurate expressions for the transport coefficients, and a first-order objective is to assure this accuracy from a
careful treatment. As is shown in this paper, the price of this approach, in contrast to more phenomenological approaches, is an increasing complexity of the expressions for the transport coefficients as the system becomes more complex. In this context, we expect the theory reported here for granular binary mixtures to be quite useful for the above numerical codes.

The present results can be also applied to determine the dispersion relations for
the hydrodynamic equations linearized about
the homogeneous cooling state. Some previous results \cite[]{GMD06} based on the standard
Sonine method have shown that the resulting equations exhibit a long wavelength
instability for three of the modes. The objective now is to revisit the above problem by using the
$\alpha$-dependence of the transport coefficients obtained in this paper (if this should be done, then for consistency reasons, the linearized Burnett corrections to the cooling rate would need to be considered in the hydrodynamic equations, since $\bnabla\bcdot\boldsymbol{j}_1^{(1)}$, $\bnabla_l\mathsfbi {P}_{kl}^{(1)}$ and $\bnabla\bcdot\boldsymbol{q}^{(1)}$ are of second order in the gradients, following the analysis by \cite{BDKS98}).
Another possible open problem is to extend the
present results (which are restricted to a low-density granular mixture) to finite
densities in the framework of the Enskog kinetic theory. Given that the NS transport
coefficients have been recently obtained from the Enskog equation \cite[]{GDH07,GHD07}
by means of the standard Sonine approximation, it would be interesting to compare again
the above results with those derived from the modified first Sonine method. Finally, it
would also be interesting to devise a method to measure by means of Monte Carlo
simulations some of the NS transport coefficients of the heat flux in a granular binary
mixture. This method could be based on the method proposed by \cite{MSG07} for a single
gas where the application of a homogeneous, anisotropic external force produces heat
flux in the absence of gradients. We plan to carry out such extensions in the near
future.

\begin{acknowledgments}
This research has been supported by the Ministerio de Educaci´on y Ciencia (Spain)
through Programa Juan de la Cierva (F.V.R.) and Grants Nos. FIS2007--60977 (V.G. and
F.V.R) and DPI2007-63559 (J.M.M.). Partial support from the Junta de Extremadura
through Grants Nos. GRU07046 (V.G. and F.V.R) and GRU07003 (J.M.M.) is also
acknowledged.
\end{acknowledgments}

\appendix
\section{Homogeneous cooling state}
\label{appB}

In this Appendix, the expressions of the cooling rate $\zeta^{(0)}$ and the
fourth-degree cumulants $c_i$ are given. These expressions were reported by
\cite{GD99b} for inelastic hard spheres ($d=3$). Here, we extend these expressions to
an arbitrary number of dimensions $d$.

By using the leading Sonine approximation (\ref{3.1}) for $f_1^{(0)}$ and neglecting nonlinear terms in $c_1$
and $c_2$, the (reduced) cooling rate $\zeta_1^*\equiv \zeta_1^{(0)}/\nu_0$ (where
$\nu_0=n\sigma_{12}^{d-1}v_0$) can be written as
\begin{equation}
\label{b1} \zeta_1^*=\zeta_{10}+\zeta_{11}c_1+\zeta_{12}c_2,
\end{equation}
where
\begin{eqnarray}
\label{b2} \zeta_{10}&=&\frac{\sqrt{2}\pi^{(d-1)/2}}{d\Gamma\left(\frac{d}{2}\right)}x_1 \left(\frac{\sigma
_1}{\sigma_{12}}\right)^{d-1}\theta_1^{-1/2}
(1-\alpha_{11}^2)+\frac{4\pi^{(d-1)/2}}{d\Gamma\left(\frac{d}{2}\right)}x_2\mu_{21}\nonumber\\
& & \times \left(\frac{1+\theta}{\theta}\right)^{1/2}(1+\alpha_{12}) \theta_2^{-1/2}
\left[1-\frac{1}{2}\mu_{21}(1+\alpha_{12})(1+\theta) \right],
\end{eqnarray}
\begin{eqnarray}
\label{b3} \zeta_{11}&=&\frac{3\pi^{(d-1)/2}}{16\sqrt{2}d\Gamma\left(\frac{d}{2}\right)} x_1 \left(\frac{\sigma
_1}{\sigma_{12}}\right)^{d-1} \theta_1^{-1/2}
(1-\alpha_{11}^2)\nonumber\\
& & +\frac{\pi^{(d-1)/2}}{8d\Gamma\left(\frac{d}{2}\right)}x_2\mu_{21}
\frac{(1+\theta)^{-3/2}}{\theta^{1/2}}(1+\alpha_{12}) \theta_2^{-1/2}\times \nonumber \\
& &
\left[2(3+4\theta)-3\mu_{21}(1+\alpha_{12})(1+\theta) \right],
\end{eqnarray}
\begin{equation}
\label{b4} \zeta_{12}=-\frac{\pi^{(d-1)/2}}{8d\Gamma\left(\frac{d}{2}\right)}x_2
\mu_{21}\left(\frac{1+\theta}{\theta}\right)^{-3/2}(1+\alpha_{12}) \theta_2^{-1/2}
\left[2+3\mu_{21}(1+\alpha_{12})(1+\theta) \right].
\end{equation}
In the above equations, $\mu_{ij}=m_i/(m_i+m_j)$, $\theta_1=1/(\mu_{21}\gamma_1)$,
$\theta_2=1/(\mu_{12}\gamma_2)$, $\theta=\theta_1/\theta_2$, and $v_0=\sqrt{2T(m_1+m_2)/m_1m_2}$. The expression
for $\zeta_2^*$ can be easily inferred from\ (\ref{b2})--(\ref{b4}) by interchanging 1 and 2 and setting
$\theta\to \theta^{-1}$.

The coefficients $c_1$ and $c_2$ are determined from the Boltzmann equations by multiplying them by $V^4$, and
integrating over the velocity. After some algebra, when only linear terms in $c_1$ and $c_2$ are retained, the
result is
\begin{equation}
\label{b5} c_1=\frac{AG-ED}{BG-DF},\quad c_2=\frac{BE-AF}{BG-DF},
\end{equation}
where
\begin{equation}
\label{b6} A=-\frac{d(d+2)}{2\theta_1^2}\zeta_{10}-\Lambda_1,\quad
E=-\frac{d(d+2)}{2\theta_2^2}\zeta_{20}-\Lambda_2,
\end{equation}
\begin{equation}
\label{b7} D=\frac{d(d+2)}{2\theta_1^2}\zeta_{12}+\Lambda_{12},\quad
F=\frac{d(d+2)}{2\theta_2^2}\zeta_{21}+\Lambda_{21},
\end{equation}
\begin{equation}
\label{b8} B=\frac{d(d+2)}{2\theta_1^2}\left(\zeta_{11}+\frac{1}{2} \zeta_{10}\right)+\Lambda_{11},\quad
G=\frac{d(d+2)}{2\theta_2^2}\left(\zeta_{22}+\frac{1}{2} \zeta_{20}\right)+\Lambda_{22}.
\end{equation}
In the above equations, $\Lambda_1$, $\Lambda_{11}$ and $\Lambda_{12}$ are given by
\begin{eqnarray}
\label{b9} \Lambda _{1} &=&-\frac{\pi^{(d-1)/2}}{\sqrt{2}\Gamma\left(\frac{d}{2}\right)} \theta _{1}^{-5/2}
 x_{1}\left(\frac{\sigma_{1}}{{\sigma}_{12}}\right)^{d-1} \frac{ 3+2d+2\alpha _{11}^{2}}{2}
\left(1-\alpha
_{11}^{2}\right)\nonumber\\
& & +\frac{\pi^{(d-1)/2}}{\Gamma\left(\frac{d}{2}\right)} \theta _{1}^{-5/2}x_2\left( 1+\theta\right)
^{-1/2}\mu_{21}\left( 1+\alpha_{12}\right)\nonumber\\
& & \times \left\{ -2\left[d+3+(d+2)\theta\right] +\mu _{21}\left( 1+\alpha _{12}\right) \left( 1+\theta \right)
\left( 11+ d+\frac{d^2+5d+6}{d+3} \theta \right)\right.\nonumber\\
& & \left. -8\mu _{21}^{2}\left( 1+\alpha _{12}\right) ^{2}\left( 1+\theta \right) ^{2} +2\mu _{21}^{3}\left(
1+\alpha _{12}\right) ^{3}\left( 1+\theta \right) ^{3}\right\} \;,
\end{eqnarray}
\begin{eqnarray}
\label{b10} \Lambda _{11} &=&-\frac{\pi^{(d-1)/2}}{\sqrt{2}\Gamma\left(\frac{d}{2}\right)} \theta _{1}^{-5/2}
 x_{1}\left(\frac{\sigma_{1}}{{\sigma}_{12}}\right)^{d-1} \left[\frac{d-1}{2}(1+\alpha_{11})+ \frac{3}{64}
\left( 10d+39+10\alpha _{11}^{2}\right) \left( 1-\alpha
_{11}^{2}\right)\right] \nonumber \\
& & +\frac{\pi^{(d-1)/2}}{16\Gamma\left(\frac{d}{2}\right)} \theta_{1}^{-5/2}x_2\left(1+\theta\right)^{-5/2}
\mu_{21}\left(
1+\alpha _{12}\right) \nonumber\\
& & \times
 \left\{-2\left[ 45+15d+(114+39d)\theta
+(88+32d)\theta^{2}+(16+8d)\theta^{3}\right] \right.  \nonumber \\
&&+3\mu _{21}\left( 1+\alpha_{12}\right) \left( 1+\theta \right) \left[ 55+5d+9(10+d)\theta+4(8+d)\theta
^{2}\right]\nonumber\\
& & \left. -24\mu _{21}^{2}\left( 1+\alpha _{12}\right) ^{2}\left( 1+\theta \right) ^{2}\left( 5+4\theta
\right)+30\mu _{21}^{3}\left( 1+\alpha _{12}\right) ^{3}\left( 1+\theta \right) ^{3}\right\} \;,
\end{eqnarray}
\begin{eqnarray}
\label{b11} \Lambda_{12} &=&\frac{\pi^{(d-1)/2}}{16\Gamma\left(\frac{d}{2}\right)}
\theta_{1}^{-5/2}x_2\theta^2\left(1+\theta\right)^{-5/2} \mu_{21}\left(
1+\alpha_{12}\right) \nonumber\\
& & \times\left\{ 2\left[ d-1+(d+2)\theta \right] +3\mu _{21}\left( 1+\alpha
_{12}\right) \left( 1+\theta\right)\left[d-1+(d+2)\theta\right] \right.  \nonumber \\
& & \left. -24\mu _{21}^{2}\left( 1+\alpha _{12}\right)^{2}\left( 1+\theta \right) ^{2}+30\mu _{21}^{3}\left(
1+\alpha_{12}\right) ^{3}\left( 1+\theta \right) ^{3}\right\}  \;.
\end{eqnarray}
As before, the expressions for $\Lambda_2$, $\Lambda_{22}$ and $\Lambda_{21}$ are easily obtained from\
(\ref{b9})--(\ref{b11}) by changing $1\leftrightarrow 2$. In the case of a three-dimensional system ($d=3$), all the above expressions reduce to those previously obtained for hard spheres \cite[]{GD99b}

The dependence of the temperature ratio $\gamma=T_1/T_2$ on the parameters of the mixture is determined by
requiring that the partial cooling rates $\zeta_i^{(0)}$ for the partial temperatures $T_i$ must be equal, i.e.,
\begin{equation}
\label{b12} \zeta_1^{(0)}=\zeta_2^{(0)}=\zeta^{(0)}.
\end{equation}
Once equation\ (\ref{b12}) is solved, the cumulants $c_i$ are explicitly obtained by substituting $\gamma$ into\
(\ref{b5}).

\section{Expressions for the collision frequencies}
\label{appA}

The expressions of the different collision frequencies appearing in the NS transport
coefficients are explicitly given in this Appendix. For their definitions please refer to the Supplementary Material file. The collision frequency $\nu_D$
associated to the the mass flux is given by
\begin{eqnarray}
\label{a9} \nu_D&=&\frac{2\pi^{(d-1)/2}}{d\Gamma\left(\frac{d}{2}\right)}
(1+\alpha_{12})
\left(\frac{\theta_1+\theta_2}{\theta_1\theta_2}\right)^{1/2}\left\{x_2\mu_{21}\left[1+\frac{1}{16}
\frac{\theta_2(3\theta_2+4\theta_1)c_1-\theta_1^2c_2}{(\theta_1+\theta_2)^2}
\right]\right.\nonumber\\
& & + \left.x_1\mu_{12}\left[1+\frac{1}{16}
\frac{\theta_1(3\theta_1+4\theta_2)c_2-\theta_2^2c_1}{(\theta_1+\theta_2)^2}
\right]\right\}.
\end{eqnarray}
It must be noted that the expression (\ref{a9}) for
$\nu_D$ has been obtained by considering only linear terms in $c_1$ and $c_2$.

The expressions of the collision frequencies $\tau_{ij}$ corresponding to the shear
viscosity are very long and so, for the sake of clarity, we have preferred to present
them in terms of the operators $\Delta_i$ defined as
\begin{equation}
\label{a9.1} \Delta_i\equiv \theta_i^2 \frac{\partial^2}{\partial
\theta_i^2}+(d+2)\theta_i \frac{\partial}{\partial \theta_i}+\frac{d(d+2)}{4}.
\end{equation}
Using (\ref{a9.1}), the collision frequencies $\tau_{11}$ and $\tau_{12}$ are given by
\begin{eqnarray}
\label{a10} \tau_{11}&=&\frac{3\sqrt{2}\pi^{(d-1)/2}}{d(d+2)\Gamma\left(\frac{d}{2}\right)}
x_1\left(\frac{\sigma_{1}}{{\sigma}_{12}}\right)^{d-1} (1+\alpha_{11})\theta_{1}^{-1/2}
\left(1+\frac{2}{3}d-\alpha_{11}\right)\left(1+\frac{7}{32}c_1\right)\nonumber\\
&&+\frac{2\pi^{(d-1)/2}}{d(d-1)(d+2)\Gamma\left(\frac{d}{2}\right)}x_2
(1+\alpha_{12})\mu_{21}(\theta_1\theta_2)^{d/2}\theta_1^2\nonumber\\
& & \times \left[1-\frac{c_1}{4}(2-\Delta_1)+
\frac{c_2}{4}\Delta_2\right]I_{\eta}^{(11)}(\theta_1,\theta_2),
\end{eqnarray}
\begin{eqnarray}
\label{a11} \tau_{12}=\frac{2\pi^{(d-1)/2}}{d(d-1)(d+2)\Gamma\left(\frac{d}{2}\right)}x_2
& & (1+\alpha_{12})\frac{\mu_{21}^2}{\mu_{12}}(\theta_1\theta_2)^{d/2}\theta_1^2 \times
\nonumber \\
& & \left[1-\frac{c_2}{4}(2-\Delta_2)+
\frac{c_1}{4}\Delta_1\right]I_{\eta}^{(12)}(\theta_1,\theta_2),
\end{eqnarray}
where
\begin{eqnarray}
\label{a12}
I_\eta^{(11)}(\theta_1,\theta_2)&=&(\theta_1\theta_2)^{-\frac{d+1}{2}}\left\{2(d+3)(d-1)
(\mu_{12}\theta_2-\mu_{21}\theta_1)\theta_1^{-2}
\left(\theta_1+\theta_2\right)^{-1/2}\right.\nonumber\\
& & +3(d-1)\mu_{21}\left(1+\frac{2d}{3}-\alpha_{12}\right)
\theta_1^{-2}\left(\theta_1+\theta_2\right)^{1/2}\nonumber\\
 & &
\left.+\left[2d(d+1)-4\right]\theta_1^{-1}\left(\theta_1+\theta_2\right)^{-1/2}\right\}
,
\end{eqnarray}
\begin{eqnarray}
\label{a13} I_\eta^{(12)}(\theta_1,\theta_2) &=&(\theta_1\theta_2)^{-\frac{d+1}{2}}\left\{
2(d+3)(d-1)(\mu_{12}\theta_2-\mu_{21}\theta_1)\theta_2^{-2}
\left(\theta_1+\theta_2\right)^{-1/2}\right.\nonumber\\
& & +3(d-1)\mu_{21}\left(1+\frac{2d}{3}-\alpha_{12}\right)
\theta_2^{-2}\left(\theta_1+\theta_2\right)^{1/2}\nonumber\\
& &
\left.-\left[2d(d+1)-4\right]\theta_2^{-1}\left(\theta_1+\theta_2\right)^{-1/2}\right\}
.
\end{eqnarray}
The corresponding expressions for $\tau_{22}$ and $\tau_{21}$ can be easily inferred from\
(\ref{a11})--(\ref{a13}).

Let us consider now the NS transport coefficients $d_i^*$, $\ell_i^*$ and
$\lambda_i^*$ of the heat flux. The quantities $Y_i$ appearing in the expressions
(\ref{3.26})--(\ref{3.27}) for these coefficients are given by
\begin{equation}
\label{3.30} Y_1= \frac{D^*}{
x_1\gamma_1^2}\left[\omega_{12}-\zeta^{*}\left(1+\frac{c_1}{2}\right)\right]-\frac{1}{\gamma_1^2}
\left(\frac{\partial \gamma_1}{\partial x_1} \right)_{p,T}\left(1+\frac{c_1}{2}\right)-
\frac{1}{2\gamma_1} \left(\frac{\partial c_1}{\partial x_1} \right)_{p,T},
\end{equation}
\begin{equation}
\label{3.31} Y_2= -\frac{D^*}{
x_2\gamma_2^2}\left[\omega_{21}-\zeta^{*}\left(1+\frac{c_2}{2}\right)\right]-\frac{1}{\gamma_2^2}
\left(\frac{\partial \gamma_2}{\partial x_1} \right)_{p,T}\left(1+\frac{c_2}{2}\right)-
\frac{1}{2\gamma_2} \left(\frac{\partial c_2}{\partial x_1} \right)_{p,T},
\end{equation}
\begin{equation}
\label{3.32}
Y_3=\frac{D_p^*}{x_1\gamma_1^2}\left[\omega_{12}-\zeta^{*}\left(1+\frac{c_1}{2}\right)\right]
-\frac{m_1n}{\rho}\frac{c_1}{2\gamma_1^2},
\end{equation}
\begin{equation}
\label{3.33}
Y_4=-\frac{D_p^*}{x_2\gamma_2^2}\left[\omega_{21}-\zeta^{*}\left(1+\frac{c_2}{2}\right)\right]
-\frac{m_2n}{\rho}\frac{c_2}{2\gamma_2^2},
\end{equation}
\begin{equation}
\label{3.34} Y_5=
\frac{D^{\prime*}}{x_1\gamma_1^2}\left[\omega_{12}-\zeta^{*}\left(1+\frac{c_1}{2}\right)\right]
-\frac{1}{\gamma_1}\left(1+\frac{c_1}{2}\right),
\end{equation}
\begin{equation}
\label{3.35}
 Y_6= -\frac{D^{\prime*}}{x_2\gamma_2^2}\left[\omega_{21}-\zeta^{*}\left(1+\frac{c_2}{2}\right)\right]
-\frac{1}{\gamma_2}\left(1+\frac{c_2}{2}\right),
\end{equation}
where
\begin{eqnarray} \label{a14}
\nu_{11}&=&\frac{8\pi^{(d-1)/2}}{d(d+2)\Gamma\left(\frac{d}{2}\right)}
x_1\left(\frac{\sigma_{1}}{{\sigma}_{12}}\right)^{d-1}(1+\alpha_{11})(2\theta_{1})^{-1/2}
\left[\frac{d-1}{2}+\frac{3}{16}(d+8)(1-\alpha_{11})\right.\nonumber\\
& & \left.+\frac{296+217 d
-3(160+11d)\alpha_{11}}{512}c_1\right]\nonumber\\
& & +\frac{\pi^{(d-1)/2}}{d(d+2)\Gamma\left(\frac{d}{2}\right)} x_2\mu_{21}(1+\alpha_{12})
(\theta_1\theta_2)^{d/2}\theta_1^3\left\{\left[1-\frac{c_1}{4}(d+8-\Delta_1)+
\frac{c_2}{4}\Delta_2\right]\right.\nonumber\\
& & \times (\theta_1\theta_2)^{-\frac{d+3}{2}}(\theta_1+\theta_2)^{-3/2}E-(d+2)\theta_1^{-1}
\left[1-\frac{c_1}{4}(d+8-\Delta_1)+ \frac{c_2}{4}\Delta_2\right]\nonumber\\
& & \left.\times \theta_1^{-\frac{d+5}{2}}\theta_2^{-\frac{d+1}{2}}(\theta_1+\theta_2)^{-1/2}[(d+2)\theta_1+
(d+3)\theta_2]\right\}\nonumber\\
& &  -\frac{\pi^{(d-1)/2}}{d\Gamma\left(\frac{d}{2}\right)} x_2\mu_{21}(1+\alpha_{12})
(\theta_1\theta_2)^{d/2}\theta_1^2\left\{\left[1-\frac{c_1}{4}(d+8-\Delta_1)+
\frac{c_2}{4}\Delta_2\right]\right.\nonumber\\
& & \times (\theta_1\theta_2)^{-\frac{d+3}{2}}(\theta_1+\theta_2)^{-1/2}A-(d+2)\theta_1^{-1}
\left[1-\frac{c_1}{4}(d+8-\Delta_1)+ \frac{c_2}{4}\Delta_2\right]\nonumber\\
& & \left.\times \theta_1^{-\frac{d+3}{2}}\theta_2^{-\frac{d+1}{2}}(\theta_1+\theta_2)^{1/2}\right\}
\nonumber\\
& & -\frac{c_1}{2}\frac{\pi^{(d-1)/2}}{d\Gamma\left(\frac{d}{2}\right)}
x_2\mu_{21} (1+\alpha_{12})(\theta_1+\theta_2)^{-1/2}\theta_1^{1/2}
\theta_2^{-3/2}\left[A-(d+2)\theta_2-(d+1) \theta_1^{-1}\theta_2^2\right],\nonumber\\
\end{eqnarray}
\begin{eqnarray}
\label{a15} \nu_{12}&=&-\frac{\pi^{(d-1)/2}}{d(d+2)\Gamma\left(\frac{d}{2}\right)}
x_2\frac{\mu_{21}^2}{\mu_{12}}(1+\alpha_{12})
(\theta_1\theta_2)^{d/2}\theta_1^3\left\{\left[1-\frac{c_2}{4}(d+8-\Delta_2)+
\frac{c_1}{4}\Delta_1\right]\right.\nonumber\\
& & \times(\theta_1\theta_2)^{-\frac{d+3}{2}}(\theta_1+\theta_2)^{-3/2}F-(d+2)\theta_1^{-1}
\left[1-\frac{c_2}{4}(d+8-\Delta_2)+ \frac{c_1}{4}\Delta_1\right]\nonumber\\
& &\left.\times \theta_1^{-\frac{d+1}{2}}\theta_2^{-\frac{d+5}{2}}(\theta_1+\theta_2)^{-1/2}[(d+3)\theta_1+
(d+2)\theta_2]\right\}\nonumber\\
& & -\frac{\pi^{(d-1)/2}}{d\Gamma\left(\frac{d}{2}\right)}x_2\frac{\mu_{21}^2}{\mu_{12}}
(1+\alpha_{12})\theta_1^{3+\frac{d}{2}}\theta_2^{\frac{d}{2}-1}\left\{
\left[1-\frac{c_2}{4}(d+8-\Delta_2)+ \frac{c_1}{4}\Delta_1\right]\right.\nonumber\\
& & \times (\theta_1\theta_2)^{-\frac{d+3}{2}}(\theta_1+\theta_2)^{-1/2}B+(d+2)\theta_1^{-1}\nonumber\\
& & \left. \left[1-\frac{c_2}{4}(d+8-\Delta_2)+ \frac{c_1}{4}\Delta_1\right]
\theta_1^{-\frac{d+1}{2}}\theta_2^{-\frac{d+3}{2}}(\theta_1+\theta_2)^{1/2} \right\} \nonumber\\
& &-\frac{\pi^{(d-1)/2}}{2d\Gamma\left(\frac{d}{2}\right)}x_2\frac{\mu_{21}^2}{\mu_{12}}(1+\alpha_{12})
(\theta_1+\theta_2)^{-1/2}\theta_1^{3/2}\theta_2^{-5/2}\left[c_2
B+c_2(d+2)(\theta_1+\theta_2)-c_1\theta_2\right],\nonumber\\
\end{eqnarray}
\begin{eqnarray}
\label{a16} \omega_{12}&=&\frac{\sqrt{2}\pi^{(d-1)/2}}{d\Gamma\left(\frac{d}{2}\right)}
x_1\left(\frac{\sigma_{1}}{{\sigma}_{12}}\right)^{d-1}\theta_1^{-1/2}(1+\alpha_{11})
\left[1-\alpha_{11}+\frac{70+47d-3(34+5d)\alpha_{11}}{32(d+2)}c_1\right]
\nonumber\\
& &
+\frac{2\pi^{(d-1)/2}}{d(d+2)\Gamma\left(\frac{d}{2}\right)}\mu_{21}(1+\alpha_{12})\theta_1^2(\theta_1\theta_2)^{d/2} \times \nonumber
\\ & &
\left[x_2 \left(1+\frac{c_1}{4}\Delta_1+\frac{c_2}{4}\Delta_2\right)(\theta_1\theta_2)^{-\frac{d+3}{2}}
(\theta_1+\theta_2)^{-1/2}A\right.\nonumber\\
& & \left. -x_1\frac{\mu_{12}}{\mu_{21}}\theta_2\theta_1^{-1}
\left(1+\frac{c_1}{4}\Delta_1+\frac{c_2}{4}\Delta_2\right)(\theta_1\theta_2)^{-\frac{d+3}{2}}
(\theta_1+\theta_2)^{-1/2}B\right.
\nonumber\\
& &-(d+2)\theta_1^{-1}x_2\left(1+\frac{c_1}{4}\Delta_1+
\frac{c_2}{4}\Delta_2\right)\theta_1^{-\frac{d+3}{2}}\theta_2^{-\frac{d+1}{2}}(\theta_1+\theta_2)^{1/2}\nonumber\\
& & \left.-(d+2)x_1\frac{\mu_{12}}{\mu_{21}}\theta_2\theta_1^{-2}\left(1+\frac{c_1}{4}\Delta_1+
\frac{c_2}{4}\Delta_2\right)\theta_1^{-\frac{d+1}{2}}\theta_2^{-\frac{d+3}{2}}
(\theta_1+\theta_2)^{1/2}\right]\nonumber\\
 & & -c_1\frac{\pi^{(d-1)/2}}{d\Gamma\left(\frac{d}{2}\right)}(1+\alpha_{12})
\left(\frac{\theta_1+\theta_2}{\theta_1\theta_2}\right)^{1/2}\left(x_2\mu_{21}+ x_1\mu_{12}\right).
\end{eqnarray}
In the above equations we have introduced the quantities
\begin{eqnarray}
\label{a17}
A(\theta_1,\theta_2)&=&(d+2)(2\beta_{12}+\theta_2)+\mu_{21}(\theta_1+\theta_2)\times
\\
& &\left\{(d+2)(1-\alpha_{12})
-[(11+d)\alpha_{12}-5d-7]\beta_{12}\theta_1^{-1}\right\}\nonumber\\
& & +3(d+3)\beta_{12}^2\theta_1^{-1}+2\mu_{21}^2\left(2\alpha_{12}^{2}-\frac{d+3}{2}\alpha
_{12}+d+1\right)\theta_1^{-1}(\theta_1+\theta_2)^2, \nonumber
\end{eqnarray}
\begin{eqnarray}
\label{a18} B(\theta_1,\theta_2)&=&
(d+2)(2\beta_{12}-\theta_1)+\mu_{21}(\theta_1+\theta_2)\times
\\& & \left\{(d+2)(1-\alpha_{12})
+[(11+d)\alpha_{12}-5d-7]\beta_{12}\theta_2^{-1}\right\}\nonumber\\
& & -3(d+3)\beta_{12}^2\theta_2^{-1}-2\mu_{21}^2\left(2\alpha_{12}^{2}-\frac{d+3}{2}\alpha
_{12}+d+1\right)\theta_2^{-1}(\theta_1+\theta_2)^2, \nonumber
\end{eqnarray}
\begin{eqnarray}
\label{a19} E(\theta_1,\theta_2)&=&
 2\mu_{21}^2\theta_1^{-2}(\theta_1+\theta_2)^2
\left(2\alpha_{12}^{2}-\frac{d+3}{2}\alpha_{12}+d+1\right)
\left[(d+2)\theta_1+(d+5)\theta_2\right]\nonumber\\
& & -\mu_{21}(\theta_1+\theta_2) \left\{\beta_{12}\theta_1^{-2}[(d+2)\theta_1+(d+5)\theta_2][(11+d)\alpha_{12}
-5d-7]\right.\nonumber\\
& & \left. -\theta_2\theta_1^{-1}[20+d(15-7\alpha_{12})+d^2(1-\alpha_{12})-28\alpha_{12}]
-(d+2)^2(1-\alpha_{12})\right\}
\nonumber\\
& & +3(d+3)\beta_{12}^2\theta_1^{-2}[(d+2)\theta_1+(d+5)\theta_2]
\nonumber\\ & &
+
2\beta_{12}\theta_1^{-1}[(d+2)^2\theta_1+(24+11d+d^2)\theta_2]\nonumber
\\& & +(d+2)\theta_2\theta_1^{-1} [(d+8)\theta_1+(d+3)\theta_2],
\end{eqnarray}
\begin{eqnarray}
\label{a20} F(\theta_1,\theta_2)&=&
 2\mu_{21}^2\theta_2^{-2}(\theta_1+\theta_2)^2
\left(2\alpha_{12}^{2}-\frac{d+3}{2}\alpha_{12}+d+1\right)
\left[(d+5)\theta_1+(d+2)\theta_2\right]\nonumber\\
& & -\mu_{21}(\theta_1+\theta_2) \left\{\beta_{12}\theta_2^{-2}[(d+5)\theta_1+(d+2)\theta_2][(11+d)\alpha_{12}
-5d-7]\right.\nonumber\\
& & \left. +\theta_1\theta_2^{-1}[20+d(15-7\alpha_{12})+d^2(1-\alpha_{12})-28\alpha_{12}]
+(d+2)^2(1-\alpha_{12})\right\}
\nonumber\\
& & +3(d+3)\beta_{12}^2\theta_2^{-2}[(d+3)\theta_1+(d+2)\theta_2]
\nonumber\\& &
-2\beta_{12}\theta_2^{-1}[(24+11d+d^2)\theta_1+(d+2)^2\theta_2]\nonumber
\\
& & +(d+2)\theta_1\theta_2^{-1} [(d+3)\theta_1+(d+8)\theta_2].
\end{eqnarray}
Here, $\beta_{12}=\mu_{12}\theta_2-\mu_{21}\theta_1$. From\ (\ref{a14})--(\ref{a20}), one easily gets the
expressions for $\omega_{21}$, $\nu_{22}$ and $\nu_{21}$ by interchanging $1\leftrightarrow 2$.

Finally, note that the expressions (\ref{a9}), (\ref{a10}), (\ref{a11}), (\ref{a14}),
(\ref{a15}), and (\ref{a16}) reduce to those previously obtained \cite[]{GM07} when one
takes Maxwellians distributions for the reference homogeneous cooling state
$f_i^{(0)}$, i.e., when $c_1=c_2=0$. Moreover, in the case of mechanically equivalent
particles, the expressions of $\nu_D$, $\tau_{ij}$ and $\nu_{ij}$ are consistent with
those recently obtained for a monocomponent gas by using the modified Sonine
approximation \cite[]{GSM07}.

\bibliographystyle{jfm}

\bibliography{jfm4revV}

\begin{thebibliography}{57}
\expandafter\ifx\csname natexlab\endcsname\relax\def\natexlab#1{#1}\fi

\bibitem[Aranson \& Tsimring(2006)]{AT06}
{\sc Aranson, I.~S. \& Tsimring, L.~V.} 2006 Patterns and collective behaviour
  in granular media: Theoretical concepts. {\em Rev. Mod. Phys.\/} {\bf 78},
  641--692.

\bibitem[Arnarson \& Willits(1998)]{AW98}
{\sc Arnarson, B. \& Willits, J.~T.} 1998 Thermal diffusion in binary mixtures
  of smooth, nearly elastic spheres with and without gravity. {\em Phys.
  Fluids.\/} {\bf 10}, 1324--1328.

\bibitem[Barrat \& Trizac(2002)]{BT02}
{\sc Barrat, A. \& Trizac, E.} 2002 Lack of energy equipartition in homogeneous
  heated binary granular mixtures. {\em Gran. Matt.\/} {\bf 4}, 57--63.

\bibitem[Bird(1994)]{B94}
{\sc Bird, G.~A.} 1994 {\em Molecular Gas Dynamics and the Direct Simulation
  Monte Carlo of Gas Flows\/}. Clarendon, Oxford.

\bibitem[Brey {\em et~al.\/}(1997)Brey, Dufty \& Santos]{BDS97}
{\sc Brey, J.~J., Dufty, J.~W. \& Santos, A.} 1997 Dissipative dynamics for
  hard spheres. {\em J. Stat. Phys.\/} {\bf 87}, 1051--1066.

\bibitem[Brey {\em et~al.\/}(1998)Brey, Dufty, Santos \& Kim]{BDKS98}
{\sc Brey, J.~J., Dufty, J.~W., Santos, A. \& Kim, C.~S.} 1998 Hydrodynamics
  for granular flows at low density. {\em Phys. Rev. E\/} {\bf 58}, 4638--4653.

\bibitem[Brey \& Ruiz-Montero(2004)]{BR04}
{\sc Brey, J.~J. \& Ruiz-Montero, M.~J.} 2004 Simulation study of the
  \textsc{G}reen-\textsc{K}ubo relations for dilute granular gases. {\em Phys.
  Rev. E\/} {\bf 70}, 051301.

\bibitem[Brey {\em et~al.\/}(1999)Brey, Ruiz-Montero \& Cubero]{BRC99}
{\sc Brey, J.~J., Ruiz-Montero, M.~J. \& Cubero, D.} 1999 On the validity of
  linear hydrodynamics for low-density granular flows described by the
  \textsc{B}oltzmann equation. {\em Europhys. Lett.\/} {\bf 48}, 359--364.

\bibitem[Brey {\em et~al.\/}(2000)Brey, Ruiz-Montero, Cubero \&
  Garc\'{\i}a-Rojo]{BRCG00}
{\sc Brey, J.~J., Ruiz-Montero, M.~J., Cubero, D. \& Garc\'{\i}a-Rojo, R.} 2000
  Self-diffusion in freely evolving granular gases. {\em Phys. Fluids.\/} {\bf
  12}, 876--883.

\bibitem[Brey {\em et~al.\/}(2005{\natexlab{{\em a\/}}})Brey, Ruiz-Montero,
  Maynar \& Garc\'ia~de Soria]{BRMG05}
{\sc Brey, J.~J., Ruiz-Montero, M.~J., Maynar, P. \& Garc\'ia~de Soria, I.}
  2005{\natexlab{{\em a\/}}} Hydrodynamic modes, \textsc{G}reen-\textsc{K}ubo
  relations, and velocity correlations in dilute granular gases. {\em J. Phys.:
  Condens. Matter\/} {\bf 17}~(S2502).

\bibitem[Brey {\em et~al.\/}(2005{\natexlab{{\em b\/}}})Brey, Ruiz-Montero \&
  Moreno]{BRM05}
{\sc Brey, J.~J., Ruiz-Montero, M.~J. \& Moreno, F.} 2005{\natexlab{{\em b\/}}}
  Energy partition and segregation for an intruder in a vibrated granular
  system under gravity. {\em Phys. Rev. Lett.\/} {\bf 95}, 098001.

\bibitem[Brey {\em et~al.\/}(2006)Brey, Ruiz-Montero \& Moreno]{BRM06}
{\sc Brey, J.~J., Ruiz-Montero, M.~J. \& Moreno, F.} 2006 Hydrodynamic profiles
  for an impurity in an open vibrated granular gas. {\em Phys. Rev. E\/} {\bf
  73}, 031301.

\bibitem[Brey {\em et~al.\/}(2002)Brey, Ruiz-Montero, Moreno \&
  Garc\'{\i}a-Rojo]{BRMG02}
{\sc Brey, J.~J., Ruiz-Montero, M.~J., Moreno, F. \& Garc\'{\i}a-Rojo, R.} 2002
  Transversal inhomogeneities in dilute vibrofluidized granular fluids. {\em
  Phys. Rev. E\/} {\bf 65}, 061302.

\bibitem[Brilliantov \& P\"oschel(2004)]{BP04}
{\sc Brilliantov, N. \& P\"oschel, T.} 2004 {\em Kinetic Theory of Granular
  Gases\/}. Oxford University Press.

\bibitem[Chapman \& Cowling(1970)]{CC70}
{\sc Chapman, S. \& Cowling, T.~G.} 1970 {\em The Mathematical Theory of
  Nonuniform Gases\/}. Cambridge University Press.

\bibitem[Clerc {\em et~al.\/}(2008)Clerc, Cordero, Dunstan, Huff, Mujica, Risso
  \& Varas]{Clerc08}
{\sc Clerc, M.~G., Cordero, P., Dunstan, J., Huff, K., Mujica, N., Risso, D. \&
  Varas, G.} 2008 Liquid-solid-like transition in quasi-one-dimensional driven
  granular media. {\em Nature Phys.\/} {\bf 4}, 249--254.

\bibitem[Dahl {\em et~al.\/}(2002)Dahl, Hrenya, Garz\'o \& Dufty]{DHGD02}
{\sc Dahl, S.~R., Hrenya, C.~M., Garz\'o, V. \& Dufty, J.~W.} 2002 Kinetic
  temperatures for a granular mixture. {\em Phys. Rev. E\/} {\bf 66}, 041301.

\bibitem[Feitosa \& Menon(2002)]{FM02}
{\sc Feitosa, K. \& Menon, N.} 2002 Breakdown of energy equipartition in a 2d
  binary vibrated granular gas. {\em Phys. Rev. Lett.\/} {\bf 88}, 198301.

\bibitem[Garz\'o(2005)]{G05}
{\sc Garz\'o, V.} 2005 Instabilities in a free granular fluid described by the
  \textsc{E}nskog equation. {\em Phys. Rev. E\/} {\bf 72}, 021106.

\bibitem[Garz\'o \& Dufty(1999{\natexlab{{\em a\/}}})]{GD99a}
{\sc Garz\'o, V. \& Dufty, J.~W.} 1999{\natexlab{{\em a\/}}} Dense fluid
  transport for inelastic hard spheres. {\em Phys. Rev. E\/} {\bf 59},
  5895--5911.

\bibitem[Garz\'o \& Dufty(1999{\natexlab{{\em b\/}}})]{GD99b}
{\sc Garz\'o, V. \& Dufty, J.~W.} 1999{\natexlab{{\em b\/}}} Homogeneous
  cooling state for a granular mixture. {\em Phys. Rev. E\/} {\bf 60},
  5706--5713.

\bibitem[Garz\'o \& Dufty(2002)]{GD02}
{\sc Garz\'o, V. \& Dufty, J.~W.} 2002 Hydrodynamics for a granular binary
  mixture at low density. {\em Phys. Fluids.\/} {\bf 14}, 1476--1490.

\bibitem[Garz\'o {\em et~al.\/}(2007{\natexlab{{\em a\/}}})Garz\'o, Dufty \&
  Hrenya]{GDH07}
{\sc Garz\'o, V., Dufty, J.~W. \& Hrenya, C.~M.} 2007{\natexlab{{\em a\/}}}
  Enskog theory for polydisperse granular mixtures. \textsc{I}.
  \textsc{N}avier-\textsc{S}tokes order transport. {\em Phys. Rev. E\/} {\bf
  76}, 031303.

\bibitem[Garz\'o {\em et~al.\/}(2007{\natexlab{{\em b\/}}})Garz\'o, Hrenya \&
  Dufty]{GHD07}
{\sc Garz\'o, V., Hrenya, C.~M. \& Dufty, J.~W.} 2007{\natexlab{{\em b\/}}}
  Enskog theory for polydisperse granular mixtures \textsc{II}. \textsc{S}onine
  polynomial approximation. {\em Phys. Rev. E\/} {\bf 76}, 031304.

\bibitem[Garz\'o \& Montanero(2002)]{GM02}
{\sc Garz\'o, V. \& Montanero, J.~M.} 2002 Transport coefficients of a heated
  granular gas. {\em Physica A\/} {\bf 313}, 336--356.

\bibitem[Garz\'o \& Montanero(2004)]{GM04}
{\sc Garz\'o, V. \& Montanero, J.~M.} 2004 Diffusion of impurities in a
  granular gas. {\em Phys. Rev. E\/} {\bf 69}, 021301.

\bibitem[Garz\'o \& Montanero(2007)]{GM07}
{\sc Garz\'o, V. \& Montanero, J.~M.} 2007 \textsc{N}avier-\textsc{S}tokes
  transport coefficients of $d$-dimensional granular binary mixtures at
  low-density. {\em J. Stat. Phys.\/} {\bf 129}, 27--58.

\bibitem[Garz\'o {\em et~al.\/}(2006)Garz\'o, Montanero \& Dufty]{GMD06}
{\sc Garz\'o, V., Montanero, J.~M. \& Dufty, J.~W.} 2006 Mass and heat fluxes
  for a binary granular mixture at low-density. {\em Phys. Fluids.\/} {\bf 18},
  083305.

\bibitem[Garz\'o \& Santos(2003)]{GS03}
{\sc Garz\'o, V. \& Santos, A.} 2003 {\em Kinetic Theory of Gases in Shear
  Flows. Nonlinear Transport\/}. Kluwer Academic Publishers.

\bibitem[Garz\'o {\em et~al.\/}(2007{\natexlab{{\em c\/}}})Garz\'o, Santos \&
  Montanero]{GSM07}
{\sc Garz\'o, V., Santos, A. \& Montanero, J.~M.} 2007{\natexlab{{\em c\/}}}
  Modified \textsc{S}onine approximation for the
  \textsc{N}avier-\textsc{S}tokes transport coefficients of a granular gas.
  {\em Physica A\/} {\bf 376}, 94--107.

\bibitem[Goldhirsch(2003)]{G03}
{\sc Goldhirsch, I.} 2003 Rapid granular flows. {\em Annu. Rev. Fluid Mech.\/}
  {\bf 22}, 57--92.

\bibitem[Goldshtein \& Shapiro(1995)]{GS95}
{\sc Goldshtein, A. \& Shapiro, M.} 1995 Mechanics of collisional motion of
  granular materials. \textsc{P}art 1. \textsc{G}eneral hydrodynamic equations.
  {\em J. Fluid Mech.\/} {\bf 282}, 75--114.

\bibitem[Hrenya {\em et~al.\/}(2008)Hrenya, Galvin \& Wildman]{HGW08}
{\sc Hrenya, C.~M., Galvin, J.~E. \& Wildman, R.~D.} 2008 Evidence of
  higher-order effects in thermally driven granular flows. {\em J. Fluid
  Mech.\/} {\bf 598}, 429--450.

\bibitem[Huan {\em et~al.\/}(2004)Huan, Yang, Candela, Mair \&
  Walsworth]{HYCMW04}
{\sc Huan, C., Yang, X., Candela, D., Mair, R.~W. \& Walsworth, R.~L.} 2004
  \textsc{NMR} experiments on a three-dimensional vibrofluidized granular
  medium. {\em Phys. Rev. E\/} {\bf 69}, 041302.

\bibitem[Jenkins \& Mancini(1989)]{JM89}
{\sc Jenkins, J.~T. \& Mancini, F.} 1989 Kinetic theory for binary mixtures of
  smooth, nearly elastic spheres. {\em Phys. Fluids. A\/} {\bf 1}, 2050--2057.

\bibitem[Krouskop \& Talbot(2003)]{KT03}
{\sc Krouskop, P. \& Talbot, J.} 2003 Mass and size effects in
  three-dimensional vibrofluidized granular mixtures. {\em Phys. Rev. E\/} {\bf
  68}, 021304.

\bibitem[Lois {\em et~al.\/}(2007)Lois, Lema\^itre \& Carlson]{LLC07}
{\sc Lois, G., Lema\^itre, A. \& Carlson, J.~M.} 2007 Spatial force
  correlations in granular shear flow. ii. \textsc{T}heoretical implications.
  {\em Phys. Rev. E\/} {\bf 76}, 021303.

\bibitem[Lutsko(2005)]{L05}
{\sc Lutsko, J.} 2005 Transport properties of dense dissipative hard-sphere
  fluids for arbitrary energy loss models. {\em Phys. Rev. E\/} {\bf 72},
  021306.

\bibitem[Lutsko {\em et~al.\/}(2002)Lutsko, Brey \& Dufty]{LBD02}
{\sc Lutsko, J., Brey, J.~J. \& Dufty, J.~W.} 2002 Diffusion in a granular
  fluid. \textsc{II}. \textsc{S}imulation. {\em Phys. Rev. E\/} {\bf 65},
  051304.

\bibitem[Montanero \& Garz\'o(2002)]{MG02}
{\sc Montanero, J.~M. \& Garz\'o, V.} 2002 Monte \textsc{C}arlo simulation of
  the homogeneous cooling state for a granular mixture. {\em Gran. Matt.\/}
  {\bf 4}, 17--24.

\bibitem[Montanero \& Garz\'o(2003)]{MG03}
{\sc Montanero, J.~M. \& Garz\'o, V.} 2003 Shear viscosity for a heated
  granular binary mixture at low-density. {\em Phys. Rev. E\/} {\bf 67},
  021308.

\bibitem[Montanero {\em et~al.\/}(2005)Montanero, Santos \& Garz\'o]{MSG05}
{\sc Montanero, J.~M., Santos, A. \& Garz\'o, V.} 2005 \textsc{DSMC} evaluation
  of the \textsc{N}avier-\textsc{S}tokes shear viscosity of a granular fluid.
  In {\em Rarefied Gas Dynamics 24\/} (ed. AIP~Conference Proceedings), , vol.
  762, pp. 797--802.

\bibitem[Montanero {\em et~al.\/}(2007)Montanero, Santos \& Garz\'o]{MSG07}
{\sc Montanero, J.~M., Santos, A. \& Garz\'o, V.} 2007 First-order
  \textsc{C}hapman-\textsc{E}nskog velocity distribution function in a granular
  gas. {\em Physica A\/} {\bf 376}, 75--93.

\bibitem[Noskowicz {\em et~al.\/}(2007)Noskowicz, Bar-Lev, Serero \&
  Goldhirsch]{NBSG07}
{\sc Noskowicz, S.~H., Bar-Lev, O., Serero, D. \& Goldhirsch, I.} 2007
  Computer-aided kinetic theory and granular gases. {\em Europhys. Lett.\/}
  {\bf 79}, 60001.

\bibitem[Pagnani {\em et~al.\/}(2002)Pagnani, Marconi \& Puglisi]{PMP02}
{\sc Pagnani, R., Marconi, U. M.~B. \& Puglisi, A.} 2002 Driven low density
  granular mixtures. {\em Phys. Rev. E\/} {\bf 66}, 051304.

\bibitem[Rericha {\em et~al.\/}(2002)Rericha, Bizon, Shattuck \&
  Swinney]{RBSS02}
{\sc Rericha, E.~C., Bizon, C., Shattuck, M.~D. \& Swinney, H.~L.} 2002 Shocks
  in supersonic sand. {\em Phys. Rev. Lett.\/} {\bf 88}, 014302.

\bibitem[Santos {\em et~al.\/}(2004)Santos, Garz\'o \& Dufty]{SGD04}
{\sc Santos, A., Garz\'o, V. \& Dufty, J.~W.} 2004 Inherent rheology of a
  granular fluid in uniform shear flow. {\em Phys. Rev. E\/} {\bf 69}, 061303.

\bibitem[Schr\"oter {\em et~al.\/}(2006)Schr\"oter, Ulrich, Kreft, Swift \&
  Swinney]{SUKSS06}
{\sc Schr\"oter, M., Ulrich, S., Kreft, J., Swift, J.~B. \& Swinney, H.~L.}
  2006 Mechanisms in the size segregation of a binary granular mixture. {\em
  Phys. Rev. E\/} {\bf 74}, 011307.

\bibitem[Sela \& Goldhirsch(1998)]{SG98}
{\sc Sela, N. \& Goldhirsch, I.} 1998 Hydrodynamic equations for rapid flows of
  smooth inelastic spheres to \textsc{B}urnett order. {\em J. Fluid Mech.\/}
  {\bf 361}, 41--74.

\bibitem[Serero {\em et~al.\/}(2006)Serero, Goldhirsch, Noskowicz \&
  Tan]{SGNT06}
{\sc Serero, D., Goldhirsch, I., Noskowicz, S.~H. \& Tan, M.~L.} 2006
  Hydrodynamics of granular gases and granular gas mixtures. {\em J. Fluid
  Mech.\/} {\bf 554}, 237--258.

\bibitem[Vega~Reyes \& Urbach(2008{\natexlab{{\em a\/}}})]{VU08}
{\sc Vega~Reyes, F. \& Urbach, J.~S.} 2008{\natexlab{{\em a\/}}} The effect of
  inelasticity on the phase transitions of a thin vibrated granular layer. {\em
  Phys. Rev. E\/} {\bf 78}, 051301.

\bibitem[Vega~Reyes \& Urbach(2008{\natexlab{{\em b\/}}})]{VU07}
{\sc Vega~Reyes, F. \& Urbach, J.~S.} 2008{\natexlab{{\em b\/}}} Steady base
  states for \textsc{N}avier-\textsc{S}tokes granulodynamics with boundary
  heating and shear. {\em J. Fluid. Mech.\/} (submitted). Preprint:
  ar\textsc{X}iv: 0807.5125.

\bibitem[Wang {\em et~al.\/}(2003)Wang, Jin \& Ma]{WJM03}
{\sc Wang, H., Jin, G. \& Ma, Y.} 2003 Simulation study on kinetic temperatures
  of vibrated binary granular mixtures. {\em Phys. Rev. E\/} {\bf 68}, 031301.

\bibitem[Wildman \& Parker(2002)]{WP02}
{\sc Wildman, R.~D. \& Parker, D.~J.} 2002 Coexistence of two granular
  temperatures in binary vibrofluidized beds. {\em Phys. Rev. Lett.\/} {\bf
  88}, 064301.

\bibitem[Willits \& Arnarson(1999)]{WA99}
{\sc Willits, J.~T. \& Arnarson, B.} 1999 Kinetic theory of a binary mixture of
  nearly elastic disks. {\em Phys. Fluids.\/} {\bf 11}, 3116--3122.

\bibitem[Yang {\em et~al.\/}(2002)Yang, Huan, Candela, Mair \&
  Walsworth]{YHCMW02}
{\sc Yang, X., Huan, C., Candela, D., Mair, R.~W. \& Walsworth, R.~L.} 2002
  Measurements of grain motion in a dense, three-dimensional granular fluid.
  {\em Phys. Rev. Lett.\/} {\bf 88}, 044301.

\bibitem[Zamankhan(1995)]{Z95}
{\sc Zamankhan, Z.} 1995 Kinetic theory for multicomponent dense mixtures of
  slightly inelastic spherical particles. {\em Phys. Rev. E\/} {\bf 52},
  4877--4891.

\end{thebibliography}

\end{document}